\documentclass[journal]{IEEEtran}
\makeatletter
\def\endthebibliography{%
	\def\@noitemerr{\@latex@warning{Empty `thebibliography' environment}}%
	\endlist
}
\makeatother

\usepackage{cite}

\usepackage[pdftex]{graphicx} 

\usepackage[caption=false,font=footnotesize]{subfig}
\usepackage[export]{adjustbox}

\usepackage{amsmath}
\usepackage{amsfonts}
\usepackage{bm}
\usepackage{filecontents}
\usepackage{amssymb}
\usepackage{color}

\usepackage{epsfig}

\usepackage{verbatim}

\usepackage{url}

\usepackage{algorithm, tabularx}

\usepackage{lettrine}

\usepackage{lipsum}

\usepackage{siunitx}

\usepackage{soul,color}

\usepackage{mathrsfs}

\usepackage{array}

\usepackage[inline]{enumitem}

\usepackage{dblfloatfix} 

\usepackage{algorithm}

\usepackage{algpseudocode}

\makeatletter
\DeclareRobustCommand{\iscircle}{\mathord{\mathpalette\is@circle\relax}}
\newcommand\is@circle[2]{%
  \begingroup
  \sbox\z@{\raisebox{\depth}{$\m@th#1\bigcirc$}}%
  \sbox\tw@{$#1\square$}%
  \resizebox{!}{\ht\tw@}{\usebox{\z@}}%
  \endgroup
}
\makeatother

\usepackage{hhline}

\usepackage{setspace}
\usepackage{multirow}
\usepackage{array}
\newcolumntype{L}[1]{>{\raggedright\let\newline\\\arraybackslash\hspace{0pt}}m{#1}}
\newcolumntype{C}[1]{>{\centering\let\newline\\\arraybackslash\hspace{0pt}}m{#1}}
\newcolumntype{R}[1]{>{\raggedleft\let\newline\\\arraybackslash\hspace{0pt}}m{#1}}

\begin{document}
	
%\title{Channel2World: A Wireless Foundation Model for \\RF Environment Representation}
%\title{{\sffamily\bfseries Channel2World}: A Wireless Foundation Model for RF Environment Representation}

%\title{{\sffamily\bfseries\scshape Channel2World}: A Wireless Foundation Model for RF Environment Representation}

\title{{\scshape Channel2World}: A Wireless Foundation Model for RF Environment Representation}

%\title{{\fontfamily{phv}\selectfont\bfseries Channel2World}: A Wireless Foundation Model for RF Environment Representation}

%\title{{\fontfamily{phv}\selectfont\bfseries Channel2World}: A Wireless Foundation Model for RF Environments from MIMO Channels}

%\title{{\fontfamily{phv}\selectfont\bfseries Channel2World}: A Wireless Foundation Model for Learning RF Environment Embeddings from MIMO Channels}

%\title{{\textsf{\textbf{Channel2World}}}: A Wireless Foundation Model for RF Environment Representation}

%A Wireless Foundation Model for \\Environment Representation

%A Wireless Foundation Model for World Embedding

%Channel2World: A Wireless Foundation Model for \\World Embedding

%Channel2World: A Foundation Model for \\Wireless Environment Representation

%CORAL: A Wireless Foundation Model for Channel Environment Representation Learning

%Wireless Foundation Model for \\Channel Environment Representation

	\author{Hyung-Joo Moon,~\IEEEmembership{Member,~IEEE}, Joonkyu Jang,~\IEEEmembership{Graduate Student Member,~IEEE},\\ Kwang Soon Kim,~\IEEEmembership{Senior Member,~IEEE}, Seong-Lyun Kim,~\IEEEmembership{Senior Member,~IEEE},\\ Robert W. Heath, Jr.,~\IEEEmembership{Fellow,~IEEE}, and Chan-Byoung Chae,~\IEEEmembership{Fellow,~IEEE}

		\thanks{H.-J. Moon, J. Jang, and C.-B. Chae are with the School of Integrated Technology, Yonsei University, Seoul 03722, South Korea (e-mail: \{moonhj, joonkyuj, cbchae\}@yonsei.ac.kr).  K. S. Kim and S.-L. Kim are with the School of Electrical and Electronic Engineering, Yonsei University, Seoul 03722, South Korea (e-mail: \{ks.kim, slkim\}@yonsei.ac.kr). R. W. Heath, Jr. is with the Department of Electrical and Computer Engineering, University of California, San Diego, La Jolla, CA 92093, USA (e-mail: rwheathjr@ucsd.edu). \emph{Corresponding author: Chan-Byoung Chae}.}
        }
	
	\maketitle

\begin{abstract}
Wireless channels are commonly treated as link-specific observations, although their multipath structure is governed by the surrounding radio-frequency (RF) environment. In this paper, we propose Channel2World, a wireless foundation model that learns a reusable environment-level representation from multiple-input multiple-output (MIMO) channel-position observations. The model aggregates channels collected within the same base-station-centered environment into a \emph{wireless world embedding} using a Transformer-based encoder. The encoder is pretrained through context-query prediction, where context channels condition user equipment (UE) position and relative path-gain prediction for disjoint query channels. After pretraining, the encoder is frozen and used as a task-agnostic environment-conditioning module for downstream wireless models, enabling adaptation to unseen environments without site-specific fine-tuning. To learn an environment-level latent space that generalizes across deployments, we pretrain Channel2World using ray-tracing data from 26,000 environments, with approximately 5,000 channel measurements per environment. Evaluations on UE localization, beam-domain channel state information (CSI) reconstruction, and RF-observable geometry reconstruction show that the learned embeddings provide effective conditioning in unseen environments. For localization and CSI reconstruction tasks, embedding-based conditioning outperforms or remains competitive with site-specific fine-tuning, although fine-tuning requires task-specific labeled data and additional gradient-based adaptation. The embeddings also support the reconstruction of dominant reflector structures, indicating their utility as reusable environmental priors across tasks.
\end{abstract}


	\IEEEpeerreviewmaketitle

\begin{figure*}[t]
	\begin{center}
		{\includegraphics[width=2.0\columnwidth,keepaspectratio]
			{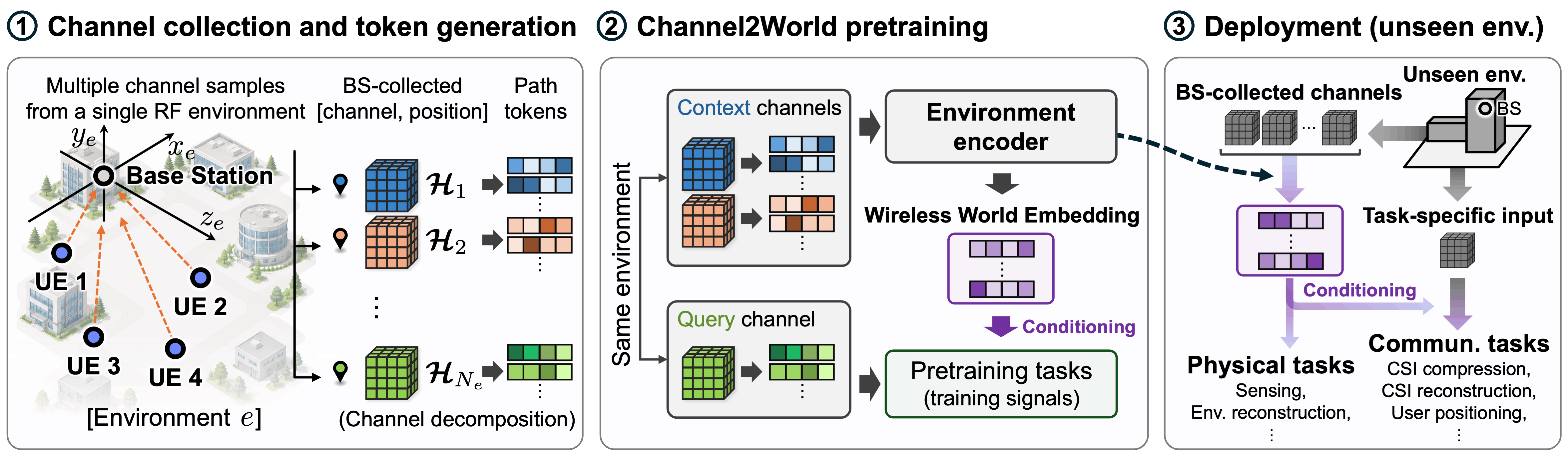}
            \caption{Multiple BS-collected MIMO channel-position observations from the same RF environment are encoded into a wireless world embedding, which provides environment-level conditioning for downstream wireless and RF-observable inference tasks.}
			\label{fig:sys}
		}
	\end{center}
	\vspace{-10pt}
\end{figure*}

\section{Introduction}

\IEEEPARstart{M}{assive} multiple-input multiple-output (MIMO) and high-frequency communication systems have expanded the role of wireless channel data. With large antenna arrays, wide bandwidths, and directional propagation, a channel observation contains structured information across the spatial, delay, and frequency domains. The multipath components of a wireless channel are determined by the base-station (BS) position, the user equipment (UE) position, surrounding reflectors and scatterers, and material-dependent propagation effects~\cite{hoydis2023sionna}. A MIMO channel is therefore not only a communication variable to be estimated for link adaptation or beamforming, but also a partial radio-frequency (RF) observation of the physical propagation environment.

This interpretation gives wireless channel data a broader role in artificial intelligence (AI)-native wireless systems. Beyond serving as an input or target for link-level inference, channel-position observations can act as RF sensing measurements of the deployment environment. Existing learning-based wireless techniques have primarily used each channel observation as an individual sample for a specific task, such as channel estimation, channel state information (CSI) feedback, beam prediction, localization, or resource allocation~\cite{wen2018csinet,he2019modeldriven,heng2021sitespecific}. Recent wireless foundation models extend this paradigm by pretraining reusable models from large datasets of channels, CSI sequences, channel impulse responses (CIRs), or in-phase and quadrature (I/Q) signals~\cite{alikhani2024lwm,liu2025wifo,yang2025wirelessgpt,guler2025wimae,jiang2025csiclip,mashaal2025iqfm}. In this work, a wireless foundation model refers to a model that is pretrained on large-scale wireless observations and then reused as a common backbone or conditioning module for downstream wireless tasks. This paper considers a complementary form of wireless foundation modeling, in which the representation target is not an individual wireless observation but the RF propagation environment jointly observed through many position-indexed channels. Such an environment-level representation can provide deployment-specific propagation context to downstream models.

This context is important because environment dependence is a pervasive challenge in wireless learning. For a given task, the relationship between the available observation and the desired output can vary across environments, as the surrounding geometry determines reflector structure and path visibility. Existing methods address this dependence in two ways. One strategy is to collect site-specific data and train or fine-tune a task model for each deployment~\cite{moon2026fr3beam,aboulfotouh2025wavesfm}. This strategy can be effective, but it requires task-labeled target-site data and may become unreliable when such samples are scarce. A second strategy is to provide explicit environmental context, such as a map, point cloud, image, radio map, or neural scene representation~\cite{zheng2025musefm}. Such context can support generalization across environments, but acquiring, calibrating, and updating it can be difficult in practical deployments.

In this paper, we infer a deployment-specific environment prior directly from wireless observations. A single channel measured at one UE position provides only a local RF observation of the environment. In contrast, channel-position samples collected across one site provide multiple RF observations of the same underlying propagation structure. We therefore train an environment encoder that maps these observations to a reusable prior capturing their shared propagation structure. This approach avoids the need for explicit scene information or task-labeled target-site samples.

Based on this concept, we propose {\sffamily\scshape Channel2World}, a wireless foundation model that infers a BS-centered RF environment representation from channel-position observations. We refer to this inferred environment-level representation as a \emph{wireless world embedding}. As illustrated in Fig.~\ref{fig:sys}, multiple BS-collected channel-position samples from the same RF environment are decomposed into path tokens and aggregated by the environment encoder. During pretraining, context channels are used to infer the wireless world embedding, while disjoint query channels from the same environment provide supervision through position and relative path-gain prediction. After pretraining, the environment encoder is frozen and reused for downstream tasks. For a new environment, a modest context set of channel-position observations is used to obtain the wireless world embedding, which then conditions downstream models.

%This differs from site-specific training or fine-tuning, which typically requires a larger set of task-labeled target-site samples and an additional gradient-based adaptation process for each deployment.

\subsection{Related Work and Our Contributions}
\label{subsec:prior_work}

Deep learning for wireless communications was first developed mainly as a task-specific physical-layer tool. Neural models have been used for CSI feedback~\cite{wen2018csinet}, channel estimation and signal detection~\cite{ye2018power}, beam prediction and beam management~\cite{alrabeiah2020deep,moon2026fr3beam}, and end-to-end transceiver design~\cite{oshea2017introduction}. These studies show that wireless channels and signals contain learnable structure that can improve physical-layer processing. At the same time, the trained models are often tied to a particular task, antenna configuration, channel distribution, or deployment site. This limitation affects both task generalization and environment generalization, since a model trained for one physical-layer objective or one deployment environment may not transfer reliably to another~\cite{heng2021sitespecific,moon2026fr3beam}.

Wireless foundation models address task dependence by pretraining reusable backbones from large-scale wireless data~\cite{jiang2025cfm}. Existing models learn channel embeddings from measured or simulated channels~\cite{alikhani2024lwm}, perform masked prediction over space, time, and frequency~\cite{liu2025wifo}, use generative pretraining for wireless communication and sensing tasks~\cite{yang2025wirelessgpt}, combine masked reconstruction and contrastive learning for channel representation~\cite{guler2025wimae}, align CSI and CIR representations~\cite{jiang2025csiclip}, or extend pretraining to I/Q streams~\cite{mashaal2025iqfm}. WavesFM further demonstrates a shared wireless backbone for sensing, communication, and localization, with task-specific prediction heads and parameter-efficient fine-tuning for downstream adaptation~\cite{aboulfotouh2025wavesfm}. These studies provide an important path toward task-general wireless models. The remaining difficulty is environment dependence, since a change in local geometry, blockage, or reflector visibility can alter the relation between a wireless observation and its target. Existing foundation-model approaches therefore rely on sufficient environmental diversity during pretraining or target-site adaptation, such as parameter-efficient fine-tuning. Thus, many existing wireless foundation models reduce task dependence, while environment adaptation remains a separate site-specific process.

A more direct way to address environment dependence is to provide environmental context to wireless models. Channel knowledge maps (CKMs) store or learn location-dependent channel knowledge for environment-aware communications~\cite{zeng2021ckm,zeng2024ckmtutorial}, and CKM information has been used for hybrid beamforming design~\cite{wu2024ckmbeamforming}. Multimodal methods use light detection and ranging (LiDAR), vision, semantic information, or multi-sensor datasets for beam selection, beam alignment, and blockage prediction~\cite{klautau2019lidar,alrabeiah2020viwi,xu2023vision,yang2023environmentsemantics,alkhateeb2023deepsense}. Wireless digital twins and neural propagation models further estimate radio maps, electromagnetic fields, or site-specific channel renderers from environmental layouts or measurements~\cite{levie2021radiounet,jiang2025learnableDT,zhang2024rf3dgs,yu2024channelgpt}. Recent environment-aware foundation models, including MUSE-FM and WWM, also incorporate explicit environmental context such as scene information, point clouds, or multimodal geometric priors into wireless prediction models~\cite{zheng2025musefm,chen2026wwm}. These works show that environment priors can support environment generalization by conditioning models on deployment-specific context. However, such priors are typically supplied as an explicit scene, map, sensor observation, radio field, or site-specific neural representation. Acquiring and maintaining such priors can require additional sensing, mapping, and calibration, thereby increasing deployment overhead.

{\sffamily\scshape Channel2World} provides environment adaptation by treating BS-collected MIMO channel-position observations as a source of RF-environment sensing and encoding them into a wireless world embedding. The embedding is then used as a compact environment prior alongside task-specific inputs, allowing downstream models to condition their predictions on the deployment environment. For an unseen environment, this prior is inferred from a context set of routinely collected channel-position observations, avoiding explicit scene information and gradient-based adaptation of either the encoder or the downstream model.

The key contributions of this work are summarized as follows:
\begin{itemize}
    \item We introduce the \emph{wireless world embedding} as a compact RF environment representation for environment adaptation. The embedding is inferred from multiple channel-position observations collected in the same BS-centered environment and captures propagation structure shared across channels.

    \item We propose {\sffamily\scshape Channel2World}, a Transformer-based architecture for learning wireless world embeddings from MIMO channel observations. The model converts decomposed multipath parameters into path tokens and aggregates a variable-size context set into a fixed-size environment representation.

    \item We develop a context-query pretraining strategy that trains the environment encoder without task-specific labels. Context channels are used to infer the wireless world embedding, while disjoint query channels from the same environment provide embedding-conditioned position and path-gain prediction signals.

    \item We train and validate {\scshape Channel2World} on a ray-tracing dataset with $26000$ BS-centered propagation environments generated from randomized building layouts, with approximately $5000$ valid channel-position pairs per environment on average.

    \item We evaluate the pretrained wireless world embedding under a frozen-encoder downstream protocol. The encoder is used as an environment prior for UE position estimation from a spatial channel matrix, full discrete Fourier transform (DFT) beam-domain CSI reconstruction from sparse beam measurements, and RF-observable environment reconstruction.
\end{itemize}

\section{System Model}\label{sec:system_model}

Consider a BS that collects uplink wideband MIMO channel observations from multiple UEs over a short environment-coherence interval, during which the propagation environment is assumed to remain static. Each observation is associated with the corresponding UE position in the BS-centered coordinate system. The collected channel-position samples are therefore interpreted as multiple RF observations of the same propagation environment and are used as the measurement source for {\sffamily\scshape Channel2World}.

In this section, we first define the BS-centered coordinate system and the uplink wideband MIMO channel tensor. We then describe the multipath channel representation obtained from signal decomposition, including the angle-of-arrival (AoA), delay, and path gain of each propagation path. Although the original measurement is a wideband MIMO channel tensor, {\sffamily\scshape Channel2World} operates on the decomposed channel representation and uses path-level physical parameter estimates as its input. This representation separates the learning architecture from the number of antenna elements, array layout, and channel sounding configuration. Consequently, {\sffamily\scshape Channel2World} can be directly applied to BSs with different antenna configurations.

\subsection{Uplink Channel Tensor and Multipath Parameters}
\label{subsec:uplink_channel_tensor}

Let $e\in\{1,\cdots,E\}$ denote the BS-centered environment index. Each environment corresponds to one fixed propagation scene with a specific BS placement, antenna-array orientation, and blockage configuration, and contains $N_e$ UE channel observations. Let $\tilde{\mathbf{p}}_{e}^{\mathrm{BS}}\in\mathbb{R}^{3}$ denote the global position of the BS antenna array in environment $e$. The BS-centered coordinate system is attached to this array. The $x_e$-axis is aligned with the horizontal axis of the rectangular array, the $y_e$-axis is aligned with the vertical axis of the array, and the $z_e$-axis is the array-normal direction pointing toward the service region. Let $\mathbf{b}_{e,x},\mathbf{b}_{e,y},\mathbf{b}_{e,z}\in\mathbb{R}^{3}$ denote the orthonormal basis vectors of the BS-centered coordinate system for environment $e$, expressed in the global coordinate system. We define $\mathbf{B}_{e}=[\mathbf{b}_{e,x},\mathbf{b}_{e,y},\mathbf{b}_{e,z}]$.
For a UE with global position $\tilde{\mathbf{p}}_{e,i}^{\mathrm{UE}}$, its BS-centered position is given by
\begin{equation}
    \mathbf{p}_{e,i}
    =
    \mathbf{B}_{e}^{\mathrm{T}}
    \left(
        \tilde{\mathbf{p}}_{e,i}^{\mathrm{UE}}
        -
        \tilde{\mathbf{p}}_{e}^{\mathrm{BS}}
    \right).
    \label{eq:local_position}
\end{equation}
After this transformation, the BS array center is located at the origin, and $\mathbf{B}_{e}^{\mathrm{T}}$ maps the global BS-to-UE displacement vector to the BS-centered coordinate system. Throughout this paper, all UE positions and AoAs used by {\sffamily\scshape Channel2World} are expressed in the BS-centered coordinate system. In practice, the UE position associated with a channel observation may contain acquisition error. We denote the observed BS-centered position by
\begin{equation}
    \hat{\mathbf{p}}_{e,i}
    =
    \mathbf{p}_{e,i}
    +
    \boldsymbol{\varepsilon}_{e,i}^{\mathrm{pos}},
    \quad
    \boldsymbol{\varepsilon}_{e,i}^{\mathrm{pos}}
    \sim
    \mathcal{N}
    \left(
    \mathbf{0}_{3},
    \sigma_{\mathrm{pos}}^{2}\mathbf{I}_{3}
    \right),
    \label{eq:position_error}
\end{equation}
where $\mathbf{0}_{3}\in\mathbb{R}^{3}$ is the zero vector, $\mathbf{I}_{3}\in\mathbb{R}^{3\times3}$ is the identity matrix, and $\sigma_{\mathrm{pos}}$ denotes the position-error standard deviation per coordinate. The error is assumed independent across UE observations and symmetric across the three BS-centered coordinate axes.

The BS in environment $e$ is equipped with a rectangular antenna array of size $N_{x,e}\times N_{y,e}$, with element spacings $d_{x,e}$ and $d_{y,e}$ along the $x_e$- and $y_e$-axes, respectively, while each UE is assumed to use a single antenna. For a channel path arriving at the BS, let $\theta$ denote the angle between the arrival direction and the $z_e$-axis, and let $\phi$ denote the azimuth angle measured from the $x_e$-axis on the $x_e$-$y_e$ plane. The corresponding direction cosines along the two array axes are
\begin{equation}
    u_x(\theta,\phi)
    =
    \sin\theta\cos\phi,
    \quad
    u_y(\theta,\phi)
    =
    \sin\theta\sin\phi .
    \label{eq:direction_cosines}
\end{equation}
For an $N_{\mathrm{a}}$-element uniform linear array with spacing $d$ and carrier wavelength $\lambda$, we define the array response vector
\begin{equation}
    \mathbf{a}_{N_{\mathrm{a}},d,\lambda}(u)
    =
    [
    1,
    e^{-j2\pi d u/\lambda},
    \cdots,
    e^{-j2\pi d (N_{\mathrm{a}}-1)u/\lambda}
    ]^{\mathrm{T}},
    \label{eq:array_response_vector}
\end{equation}
where $u$ is the direction cosine along the array axis. The BS array response matrix $\mathbf{A}_{e}(\theta,\phi)\in\mathbb{C}^{N_{x,e}\times N_{y,e}}$ is then formulated as
\begin{equation}
    \mathbf{A}_{e}(\theta,\phi)
    =
    \mathbf{a}_{N_{x,e},d_{x,e},\lambda}
    \left(
    u_x(\theta,\phi)
    \right)
    \mathbf{a}_{N_{y,e},d_{y,e},\lambda}^{\mathrm{T}}
    \left(
    u_y(\theta,\phi)
    \right).
    \label{eq:array_response_matrix}
\end{equation}
The BS estimates baseband-equivalent uplink orthogonal frequency division multiplexing (OFDM) complex channels over the subcarrier set $\bold{N}_{\mathrm{sc},e}=\{0,\cdots,N_{\mathrm{sc},e}-1\}$, where $N_{\mathrm{sc},e}$ is the number of subcarriers. The baseband frequency of subcarrier $n$ is
\begin{equation}
    \nu_{e,n}
    =
    \left(
    n-\frac{N_{\mathrm{sc},e}-1}{2}
    \right)\Delta f_e ,
    \label{eq:subcarrier_frequency}
\end{equation}
where $\Delta f_e$ is the subcarrier spacing. Thus, the occupied bandwidth is $B_{\mathrm{w},e}=N_{\mathrm{sc},e}\Delta f_e$.

For UE channel $i$ in environment $e$, let $L_{e,i}$ denote the number of multipaths. For subcarrier $n\in\bold{N}_{\mathrm{sc},e}$, let $\bold{H}_{e,i,n}\in\mathbb{C}^{N_{x,e}\times N_{y,e}}$ denote the uplink spatial channel matrix across the BS array. Under a far-field finite-path model, the channel matrix is written as
\begin{equation}
\begin{aligned}
    \bold{H}_{e,i,n}
    =
    e^{j\kappa_{e,i}}
    \sum_{\ell=1}^{L_{e,i}}
    10^{\frac{p_{e,i,\ell}}{20}}
    \mathbf{A}_{e}
    (
    \theta_{e,i,\ell},
    \phi_{e,i,\ell}
    )
    e^{-j2\pi\nu_{e,n}\tau_{e,i,\ell}},
\end{aligned}
    \label{eq:tensor_channel_entry}
\end{equation}
where $\kappa_{e,i}\in[0,2\pi)$ denotes the random phase induced by the RF frontend, $p_{e,i,\ell}$ denotes the dB-scale gain of path $\ell$, $(\theta_{e,i,\ell},\phi_{e,i,\ell})$ denotes the BS-centered AoA, and $\tau_{e,i,\ell}$ denotes the path delay measured with respect to the receiver timing reference. The uplink wideband MIMO channel from UE $i$ to the BS is represented as the third-order tensor $\boldsymbol{\mathcal{H}}_{e,i}\in\mathbb{C}^{N_{x,e}\times N_{y,e}\times N_{\mathrm{sc},e}}$, obtained by stacking the subcarrier-domain spatial channel matrices along the frequency mode:
\begin{equation}
    \boldsymbol{\mathcal{H}}_{e,i}
    =
    [
    \bold{H}_{e,i,0},
    \bold{H}_{e,i,1},
    \cdots,
    \bold{H}_{e,i,N_{\mathrm{sc},e}-1}
    ],
    \label{eq:wideband_channel_tensor}
\end{equation}
where the three modes of $\boldsymbol{\mathcal{H}}_{e,i}$ correspond to the two BS array dimensions, $x_e$ and $y_e$, and the subcarrier dimension.

The delay term $\tau_{e,i,\ell}$ in~\eqref{eq:tensor_channel_entry} consists of the physical propagation delay and a timing bias due to imperfect synchronization. Let $d_{e,i,\ell}^{\mathrm{path}}$ denote the traversing distance of path $\ell$. The physical propagation delay is $\tau_{e,i,\ell}^{\mathrm{prop}}={d_{e,i,\ell}^{\mathrm{path}}}/{c_0}$,
where $c_0$ is the speed of light. The delay under the receiver timing reference is modeled as
\begin{equation}
    \tau_{e,i,\ell}
    =
    \tau_{e,i,\ell}^{\mathrm{prop}}
    +
    b_{e,i}^{\tau},
    \label{eq:biased_delay}
\end{equation}
where $b_{e,i}^{\tau}$ denotes a UE-specific timing bias that is common to all paths in the same channel observation. This bias represents the aggregate effect of receiver timing-reference selection and synchronization residuals. Furthermore, the path gain in~\eqref{eq:tensor_channel_entry} is modeled as
\begin{equation}
    p_{e,i,\ell}
    =
    p_{e,i,\ell}^{\mathrm{phy}}
    +
    \xi_{e,i,\ell}^{\mathrm{UE}}
    +
    \zeta_{e,i}^{\mathrm{g}},
    \label{eq:practical_path_power}
\end{equation}
where $p_{e,i,\ell}^{\mathrm{phy}}$ is the dB-scale path gain determined by physical propagation, $\xi_{e,i,\ell}^{\mathrm{UE}}$ is a path-dependent UE-side gain term, and $\zeta_{e,i}^{\mathrm{g}}$ is a channel-common gain-scale term. The physical propagation term $p_{e,i,\ell}^{\mathrm{phy}}$ includes distance-dependent pathloss and propagation effects determined by ray interactions, such as reflection, diffraction, scattering, and material-dependent attenuation. The UE-side term $\xi_{e,i,\ell}^{\mathrm{UE}}$ accounts for UE antenna-gain pattern, body or hand blockage, and device orientation. In the simulations, we model this term as an independent zero-mean Gaussian random variable, $\xi_{e,i,\ell}^{\mathrm{UE}}\sim \mathcal{N}(0,\sigma_\xi^2)$, where $\sigma_\xi$ denotes the standard deviation of the UE-side path-gain perturbation. The gain-scale term $\zeta_{e,i}^{\mathrm{g}}$ represents transmit-power uncertainty, receiver gain control, and calibration effects, and is common to all paths in the same channel observation.

\begin{figure*}[t]
    \centering
    \includegraphics[width=1.0\textwidth]{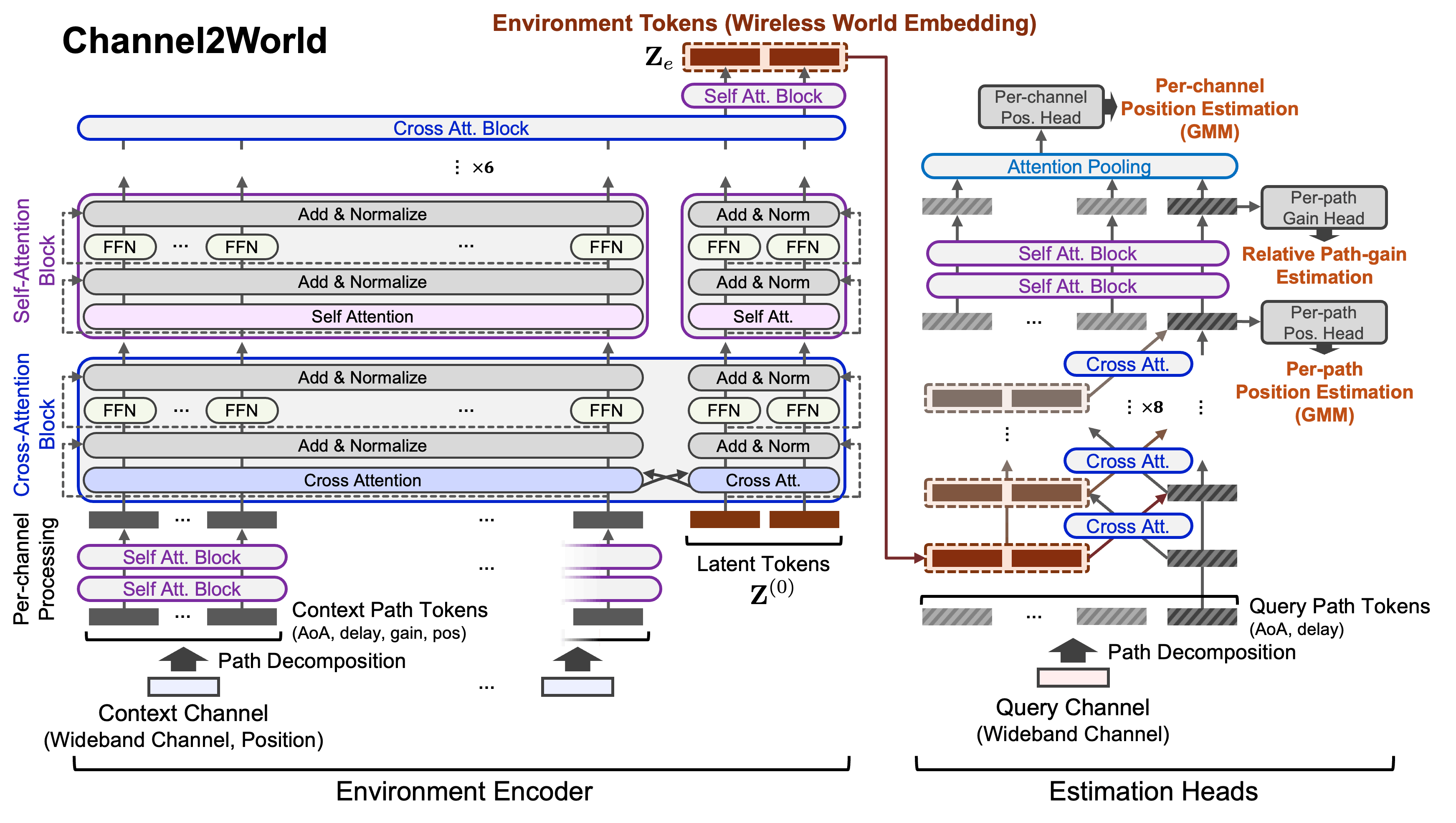}
    \caption{{\sffamily\scshape Channel2World} architecture. Path tokens from context channel-position pairs are aggregated through attention into wireless world embedding tokens, which condition query-side position and gain prediction heads.}
    \label{fig:c2w}
    \vspace{-10pt}
\end{figure*}

\subsection{Channel Decomposition and Parameter Estimation}
\label{subsec:channel_decomposition}

Before being processed by {\sffamily\scshape Channel2World}, each uplink wideband MIMO channel tensor is converted into a finite set of multipath parameters. As described in~\eqref{eq:tensor_channel_entry}, the noiseless channel tensor is represented as a sum of path-induced components with structured responses along the two spatial modes and the frequency mode. This low-rank multilinear structure supports 3D parameter estimation in the $x_e$-$y_e$-frequency domain, even in the presence of additive signal noise. Compared with 1D or 2D mixed-signal representations, the joint spatial-frequency observation provides additional separability across propagation paths, thereby significantly improving the identifiability of AoA, delay, and path-gain parameters. Existing tensor-based or subspace-based estimators can exploit this joint spatial-frequency structure to estimate these multipath parameters~\cite{zhou2017lowrank,lin2020tensor}.

We denote the pilot-based estimate of the ground-truth channel tensor $\boldsymbol{\mathcal{H}}_{e,i}$ by $\widehat{\boldsymbol{\mathcal{H}}}_{e,i}\in\mathbb{C}^{N_{x,e}\times N_{y,e}\times N_{\mathrm{sc},e}}$, which can be expressed as
\begin{equation}
    \widehat{\boldsymbol{\mathcal{H}}}_{e,i}
    =
    \boldsymbol{\mathcal{H}}_{e,i}
    +
    \boldsymbol{\mathcal{W}}_{e,i},
    \label{eq:received_tensor}
\end{equation}
where $\boldsymbol{\mathcal{W}}_{e,i}$ includes receiver noise and residual modeling errors. Using decomposition-based signal separation, the estimation of the multipath parameters is feasible from $\widehat{\boldsymbol{\mathcal{H}}}_{e,i}$, as follows:
\begin{equation}
    \big\{
    \big(
    \hat{\tau}_{e,i,\ell},
    \hat{\theta}_{e,i,\ell},
    \hat{\phi}_{e,i,\ell},
    \hat{p}_{e,i,\ell}
    \big)
    \big\}_{\ell=1}^{\hat{L}_{e,i}},
    \label{eq:raw_estimated_path_parameters}
\end{equation}
where $\hat{L}_{e,i}$ is the number of estimated paths, $\hat{\tau}_{e,i,\ell}$ is the estimated delay under the imperfect receiver timing reference, $(\hat{\theta}_{e,i,\ell},\hat{\phi}_{e,i,\ell})$ is the estimated AoA, and $\hat{p}_{e,i,\ell}$ is the estimated dB-scale path gain.

The raw estimates in~\eqref{eq:raw_estimated_path_parameters} contain the channel-common timing bias $b_{e,i}^{\tau}$ in~\eqref{eq:biased_delay} and the channel-common gain-scale factor $\zeta_{e,i}^{\mathrm{g}}$ in~\eqref{eq:practical_path_power}, reflecting the practical channel assumptions. These channel-common quantities can be removed by normalizing the estimated delays and gains within each UE channel. For the delay feature, we use the shortest estimated delay as the timing reference,
\begin{equation}
    \hat{\ell}_{e,i}^{\min}
    =
    \operatorname*{argmin}_{\ell\in\{1,\cdots,\hat{L}_{e,i}\}}
    \hat{\tau}_{e,i,\ell}.
    \label{eq:estimated_shortest_path}
\end{equation}
The relative delay is then defined as
\begin{equation}
    \hat{\bar{\tau}}_{e,i,\ell}
    =
    \hat{\tau}_{e,i,\ell}
    -
    \hat{\tau}_{e,i,\hat{\ell}_{e,i}^{\min}} .
    \label{eq:estimated_effective_delay}
\end{equation}
This normalization removes the channel-common timing bias, but also discards the absolute delay origin and preserves only the excess-delay information across paths. For the gain feature, we use the strongest estimated path as the gain reference,
\begin{equation}
    \hat{\ell}_{e,i}^{\max}
    =
    \operatorname*{argmax}_{\ell\in\{1,\cdots,\hat{L}_{e,i}\}}
    \hat{p}_{e,i,\ell}.
    \label{eq:estimated_strongest_path}
\end{equation}
The relative path gain is then defined as
\begin{equation}
    \hat{\bar{g}}_{e,i,\ell}
    =
    \hat{p}_{e,i,\ell}
    -
    \hat{p}_{e,i,\hat{\ell}_{e,i}^{\max}} .
    \label{eq:estimated_relative_gain}
\end{equation}
This normalization removes the absolute gain scale and aligns the strongest path gain to $0$~dB. After delay and gain normalization, the estimated multipath set is represented as
\begin{equation}
    \widehat{\bold{P}}_{e,i}
    =
    \{
    (
    \hat{\bar{\tau}}_{e,i,\ell},
    \hat{\theta}_{e,i,\ell},
    \hat{\phi}_{e,i,\ell},
    \hat{\bar{g}}_{e,i,\ell}
    )
    \}_{\ell=1}^{\hat{L}_{e,i}} .
    \label{eq:estimated_path_set}
\end{equation}
%Since {\sffamily\scshape Channel2World} uses the parameter-domain representation in~\eqref{eq:estimated_path_set}, its architecture is independent of the antenna-domain tensor size, array layout, and sounding configuration, provided that the extracted parameters are expressed in the BS-centered coordinate system.
In particular, the delay estimates are limited by the observation window and resolution imposed by OFDM frequency sampling. The subcarrier spacing $\Delta f_e$ determines the unambiguous delay interval, and the occupied bandwidth $B_{\mathrm{w},e}$ determines the delay resolution~\cite{zhou2017lowrank,lin2020tensor}. In the considered wideband setup with dense pilot allocation over the subcarrier domain, the unambiguous delay interval is much larger than the physical delay spread of the target environment. For example, $\Delta f_e=15$~kHz yields an unambiguous range of approximately ${c_0}/\Delta {f_e}=20$~km~\cite{mymapx}. We therefore do not model integer delay ambiguity and assume that the largest multipath propagation distance is within the unambiguous range. The finite delay resolution and other imperfections in channel decomposition are captured by the parameter-error model described below.

For each dominant path identified through multipath signal separation and associated with a ground-truth path component, the estimated parameters are modeled as
\begin{equation}
    \begin{aligned}
    \hat{\bar{\tau}}_{e,i,\ell}
    &=
    \bar{\tau}_{e,i,\ell}
    +
    \varepsilon^{\tau}_{e,i,\ell},\,
    &\hat{\theta}_{e,i,\ell}
    =
    \theta_{e,i,\ell}
    +
    \varepsilon^{\theta}_{e,i,\ell},\\
    \hat{\phi}_{e,i,\ell}
    &=
    \phi_{e,i,\ell}
    +
    \varepsilon^{\phi}_{e,i,\ell},\,
    &\hat{\bar{g}}_{e,i,\ell}
    =
    \bar{g}_{e,i,\ell}
    +
    \varepsilon^{g}_{e,i,\ell}.
    \end{aligned}
    \label{eq:path_estimation_error}
\end{equation}
The reference parameters in~\eqref{eq:path_estimation_error} are defined after removing the common delay bias and the absolute gain scale. Specifically, the ideal relative delay $\bar{\tau}_{e,i,\ell}$ and relative dB-scale path gain $\bar{g}_{e,i,\ell}$ of path $\ell$ are given by
\begin{equation}
\begin{aligned}
    \bar{\tau}_{e,i,\ell}
    &=
    \tau_{e,i,\ell}
    -
    \tau_{e,i,\ell_{e,i}^{\min}},
    \quad
    \ell_{e,i}^{\min}
    =
    \operatorname*{argmin}_{q\in\{1,\cdots,L_{e,i}\}}
    \tau_{e,i,q},\\
    \bar{g}_{e,i,\ell}
    &=
    p_{e,i,\ell}
    -
    p_{e,i,\ell_{e,i}^{\max}},
    \quad
    \ell_{e,i}^{\max}
    =
    \operatorname*{argmax}_{q\in\{1,\cdots,L_{e,i}\}}
    p_{e,i,q}.
\end{aligned}
    \label{eq:true_bias_removed_delay_gain}
\end{equation}
Here, $\ell_{e,i}^{\min}$ denotes the index of the earliest path, and $\ell_{e,i}^{\max}$ denotes the index of the strongest path. The variables $\varepsilon^{\tau}_{e,i,\ell}$, $\varepsilon^{\theta}_{e,i,\ell}$, $\varepsilon^{\phi}_{e,i,\ell}$, and $\varepsilon^{g}_{e,i,\ell}$ in~\eqref{eq:path_estimation_error} denote decomposition-equivalent estimation errors, which capture the aggregate effects of receiver noise, finite bandwidth, finite array aperture, and residual imperfections in multipath signal separation. Since channel decomposition is not the main focus of this paper, {\sffamily\scshape Channel2World} is evaluated under controlled parameter-error levels using this decomposition-equivalent error model. Specifically, the error terms are modeled as independent zero-mean Gaussian random variables: $\varepsilon^{\tau}_{e,i,\ell}\sim\mathcal{N}(0,\sigma_\tau^2)$, $\varepsilon^{\theta}_{e,i,\ell}\sim\mathcal{N}(0,\sigma_\theta^2)$, $\varepsilon^{\phi}_{e,i,\ell}\sim\mathcal{N}(0,\sigma_\phi^2)$, and $\varepsilon^{g}_{e,i,\ell}\sim\mathcal{N}(0,\sigma_g^2)$, where $\sigma_\tau$, $\sigma_\theta$, $\sigma_\phi$, and $\sigma_g$ denote the standard deviations of the relative-delay, zenith-angle, azimuth-angle, and relative-gain errors, respectively.

\section{Channel2World}
\label{sec:channel2world}

This section presents {\sffamily\scshape Channel2World}, a Transformer-based architecture that infers a wireless world embedding from channel observations collected in the same BS-centered environment. As shown in Fig.~\ref{fig:c2w}, the model consists of two parts: an environment encoder and estimation heads. The environment encoder aggregates multiple channel-position observations into latent environment tokens, without depending on the number or ordering of channel samples and multipaths. The estimation heads use the inferred environment tokens to predict target variables from query tokens that are disjoint from the context tokens, thereby providing training signals to the environment encoder.

\subsection{Environment-Level Data Organization and Token Construction}
\label{subsec:c2w_data_token}

{\sffamily\scshape Channel2World} uses the multipath parameter-domain representation, obtained from the channel decomposition described in Section~\ref{sec:system_model}. Each environment is represented by a set of channel-position pairs,
\begin{equation}
    \mathcal{D}_{e}
    =
    \{
    (
    \hat{\mathbf{p}}_{e,i},
    \widehat{\bold{P}}_{e,i}
    )
    \}_{i=1}^{N_e},
    \label{eq:c2w_environment_dataset}
\end{equation}
where $\hat{\mathbf{p}}_{e,i}$ is the observed BS-centered UE position and $\widehat{\bold{P}}_{e,i}$ is the estimated multipath set in~\eqref{eq:estimated_path_set}. Each pretraining instance is constructed within a single environment. {\sffamily\scshape Channel2World} samples a context channel-index set $\mathcal{C}_{e}$ and a query channel-index set $\mathcal{Q}_{e}$ from $\{1,\cdots,N_e\}$ such that $\mathcal{C}_{e}\cap\mathcal{Q}_{e}=\emptyset$. The context set provides the channel-position observations used to infer the wireless world embedding, whereas the query set provides disjoint channels for environment-conditioned prediction. For a channel with $\hat{L}_{e,i}$ estimated paths, we define the path-index set as $\mathcal{L}_{e,i}=\{1,\cdots,\hat{L}_{e,i}\}$. For a context channel path $\ell\in\mathcal{L}_{e,i}$ with $i\in\mathcal{C}_{e}$, the context token is defined as
\begin{equation}
\begin{aligned}
    \mathbf{x}_{e,i,\ell}^{\mathrm{ctx}}
    =
    [
    &\hat{\bar{\tau}}_{e,i,\ell},
    \sin\hat{\theta}_{e,i,\ell},
    \cos\hat{\theta}_{e,i,\ell},
    \sin\hat{\phi}_{e,i,\ell},
    \cos\hat{\phi}_{e,i,\ell},\\
    &\hat{\bar{g}}_{e,i,\ell},
    \hat{\mathbf{p}}_{e,i}^{\mathrm{T}}
    ]^{\mathrm{T}},
    \label{eq:context_token_feature}
\end{aligned}
\end{equation}
containing the relative delay, BS-centered AoA, relative path gain, and UE position corresponding to the estimated path.

In particular, query channels are selected from LoS UE channels. Let $\mathcal{I}_{e}^{\mathrm{LoS}}\subset \{1,\cdots,N_e\}$ denote the set of channel indices corresponding to UEs with an LoS path. Then, $\mathcal{Q}_{e}\subseteq\mathcal{I}_{e}^{\mathrm{LoS}}$. For these channels, the UE position provides a BS-to-UE range reference, which allows the channel timing reference to be calibrated using the LoS path. After this calibration, a propagation-delay estimate can be assigned to each path in the query channel, including non-line-of-sight (NLoS) paths. This restriction is not severe for pretraining because the query set is used only for supervision and is much smaller than the context set. For a query path $\ell\in\mathcal{L}_{e,j}$ with $j\in\mathcal{Q}_{e}$, the query token is defined as
\begin{equation}
    \mathbf{x}_{e,j,\ell}^{\mathrm{qry}}
    =
    [
    \hat{\tau}_{e,j,\ell}^{\mathrm{prop}},
    \sin\hat{\theta}_{e,j,\ell},
    \cos\hat{\theta}_{e,j,\ell},
    \sin\hat{\phi}_{e,j,\ell},
    \cos\hat{\phi}_{e,j,\ell}
    ]^{\mathrm{T}},
    \label{eq:query_token_feature}
\end{equation}
where $\hat{\tau}_{e,j,\ell}^{\mathrm{prop}}$ denotes the calibrated propagation-delay estimate of path $\ell$. The input query token does not explicitly include the observed UE position $\hat{\mathbf{p}}_{e,j}$ or the relative path gains $\{\hat{\bar{g}}_{e,j,\ell}\}_{\ell\in\mathcal{L}_{e,j}}$, which are instead used as supervised targets.

The resulting context token set and query token set are
\begin{equation}
\begin{aligned}
    \mathcal{X}_{e}^{\mathrm{ctx}}
    &=
    \{
    \mathbf{x}_{e,i,\ell}^{\mathrm{ctx}}
    \,|\,
    i\in\mathcal{C}_{e},
    \ell\in\mathcal{L}_{e,i}
    \},\\
    \mathcal{X}_{e,j}^{\mathrm{qry}}
    &=
    \{
    \mathbf{x}_{e,j,\ell}^{\mathrm{qry}}
    \,|\,
    \ell\in\mathcal{L}_{e,j}
    \},
    \quad
    j\in\mathcal{Q}_{e}.
    \label{eq:context_query_token_set}
\end{aligned}
\end{equation}
This context-query organization encourages the latent environment tokens to encode propagation information shared across channels in the same environment, allowing them to explain disjoint query channel observations that were not used to construct the embedding.

\subsection{Environment Encoder for Wireless World Embedding}
\label{subsec:c2w_environment_encoder}

The encoder maps the context token set into latent environment tokens. It follows the design of attention-based set encoders with latent tokens, where a variable-size input token set is summarized through a fixed number of trainable latent tokens using cross-attention and self-attention~\cite{lee2019settransformer,jaegle2021perceiver}. Each context token in $\mathcal{X}_{e}^{\mathrm{ctx}}$ is first embedded into a $D_{\mathrm{m}}$-dimensional feature vector by a shared linear projection. {\sffamily\scshape Channel2World} then applies self-attention within the path tokens of each context channel to encode intra-channel multipath structure before aggregating information across different channels. After channel-wise processing, the context path tokens from all channels in $\mathcal{C}_{e}$ form a variable-size set. To map this set into a fixed-size environment representation, {\sffamily\scshape Channel2World} uses $K_{\mathrm{z}}$ trainable latent tokens as aggregation slots. The input latent-token matrix is defined as
\begin{equation}
\mathbf{Z}^{(0)}
=
[
\mathbf{z}_{1}^{(0)},
\mathbf{z}_{2}^{(0)},
\cdots,
\mathbf{z}_{K_{\mathrm{z}}}^{(0)}
]^{\mathrm{T}}
\in
\mathbb{R}^{K_{\mathrm{z}}\times D_{\mathrm{m}}},
\label{eq:initial_latent_tokens}
\end{equation}
where $\mathbf{z}_{k}^{(0)}\in\mathbb{R}^{D_{\mathrm{m}}}$ is the $k$-th trainable latent token and $K_{\mathrm{z}}$ is the number of latent tokens. The same input latent-token initialization is shared across all environments and learned jointly with the other trainable model parameters.

The global aggregation stage alternates between bidirectional cross-attention and self-attention blocks. In each cross-attention block, the latent tokens attend to the processed context path tokens to collect environment information, while the context path tokens attend back to the latent tokens to incorporate information from the current latent representation. Self-attention is then applied separately within the latent-token set and the context-token set, refining the information within each token set before the next cross-attention block. Through repeated updates, the latent tokens function as fixed-size aggregation slots that summarize the variable-size context observations into the environment representation. The environment encoder output is written as
\begin{equation}
\mathbf{Z}_{e}
=
\mathrm{Enc}_{\Theta}
\big(
\mathcal{X}_{e}^{\mathrm{ctx}},
\mathbf{Z}^{(0)}
\big)
\in
\mathbb{R}^{K_{\mathrm{z}}\times D_{\mathrm{m}}},
\label{eq:environment_latent_tokens}
\end{equation}
where $\mathrm{Enc}_{\Theta}(\cdot)$ is the {\sffamily\scshape Channel2World} encoder and $\Theta$ denotes the trainable model parameters. The resulting $\mathbf{Z}_{e}$ consists of the processed latent token representations and serves as the wireless world embedding of environment $e$.

\subsection{Environment-Conditioned Estimation Heads}
\label{subsec:c2w_query_prediction}

The estimation module uses the wireless world embedding $\mathbf{Z}_{e}$ to condition predictions for query channels that are disjoint from the context channels. It contains three prediction heads: per-path position estimation, relative path-gain estimation, and per-channel position estimation. Each input query token $\mathbf{x}_{e,j,\ell}^{\mathrm{qry}}$ is first projected to dimension $D_{\mathrm{m}}$ and then interacts with $\mathbf{Z}_{e}$ through bidirectional cross-attention. This interaction allows each query path to retrieve environment information relevant to its calibrated propagation-delay estimate and AoA. The resulting environment-conditioned path token is written as
\begin{equation}
    \mathbf{r}_{e,j,\ell}^{\mathrm{path}}
    =
    \mathrm{Dec}_{\Theta}^{\mathrm{path}}
    \big(
    \mathbf{x}_{e,j,\ell}^{\mathrm{qry}},
    \mathbf{Z}_{e}
    \big)
    \in
    \mathbb{R}^{D_{\mathrm{m}}},
    \label{eq:pathwise_query_decoder}
\end{equation}
where $\mathrm{Dec}_{\Theta}^{\mathrm{path}}(\cdot)$ denotes the pathwise query-environment interaction module.

\subsubsection{Per-path position estimation}
A query path provides the BS-side ray direction through its AoA and range-related information through its propagation delay, but these quantities do not uniquely determine the UE position. The environment tokens provide the missing site-specific context by encoding reflector-related information inferred from the context channels. {\sffamily\scshape Channel2World} uses a Gaussian mixture model (GMM) head to predict a conditional density over the BS-centered UE position, $p_{\Theta}^{\mathrm{path}}(\mathbf{p}\,|\,\mathbf{r}_{e,j,\ell}^{\mathrm{path}})$, where $\mathbf{p}\in\mathbb{R}^{3}$ denotes a UE position. The GMM components are predicted from the environment-conditioned path token $\mathbf{r}_{e,j,\ell}^{\mathrm{path}}$ using a multi-layer perceptron (MLP) head with one hidden layer. During training, the predicted mixture density is evaluated at the observed query UE position $\hat{\mathbf{p}}_{e,j}$, which serves as the supervised target for each query path $\ell\in\mathcal{L}_{e,j}$.

\subsubsection{Relative path-gain estimation}
After environment-conditioned cross-attention, the path tokens are jointly refined within the query channel using intra-channel self-attention:
\begin{equation}
    \big\{
    \tilde{\mathbf{r}}_{e,j,\ell}^{\mathrm{path}}
    \big\}_{\ell\in\mathcal{L}_{e,j}}
    =
    \mathrm{Ref}_{\Theta}
    \big(
    \big\{
    \mathbf{r}_{e,j,\ell}^{\mathrm{path}}
    \big\}_{\ell\in\mathcal{L}_{e,j}}
    \big),
    \label{eq:path_refinement}
\end{equation}
where $\mathrm{Ref}_{\Theta}(\cdot)$ denotes the intra-channel self-attention block, and $\tilde{\mathbf{r}}_{e,j,\ell}^{\mathrm{path}}\in\mathbb{R}^{D_{\mathrm{m}}}$ denotes the refined path token. This step allows the prediction heads to use the relative relationships among paths observed in the same query channel. The refined path token is used to estimate the relative path gain, $\hat{\bar{g}}_{e,j,\ell}^{\mathrm{pred}}=f_{\Theta}^{\mathrm{g}}(\tilde{\mathbf{r}}_{e,j,\ell}^{\mathrm{path}})$, where $f_{\Theta}^{\mathrm{g}}:\mathbb{R}^{D_{\mathrm{m}}}\rightarrow\mathbb{R}$ is a shared gain head implemented by an MLP with one hidden layer. The supervised target for this head is the estimated relative dB-scale path gain $\hat{\bar{g}}_{e,j,\ell}$.

\subsubsection{Per-channel position estimation}
For the final channel-level position estimate, {\sffamily\scshape Channel2World} aggregates the refined path representations by attention pooling. The pooling weight of path $\ell$ in query channel $j$ is
\begin{equation}
\alpha_{e,j,\ell}
=
\frac{
\exp
\big(
\mathbf{w}_{\mathrm{p}}^{\mathrm{T}}
\tilde{\mathbf{r}}_{e,j,\ell}^{\mathrm{path}}
\big)
}{
\sum_{q\in\mathcal{L}_{e,j}}
\exp
\big(
\mathbf{w}_{\mathrm{p}}^{\mathrm{T}}
\tilde{\mathbf{r}}_{e,j,q}^{\mathrm{path}}
\big)
},
\label{eq:attention_pooling_weight}
\end{equation}
where $\mathbf{w}_{\mathrm{p}}\in\mathbb{R}^{D_{\mathrm{m}}}$ is a trainable pooling vector. The resulting channel token $\mathbf{r}_{e,j}^{\mathrm{ch}}\in\mathbb{R}^{D_{\mathrm{m}}}$ is represented as
\begin{equation}
\mathbf{r}_{e,j}^{\mathrm{ch}}
=
\sum_{\ell\in\mathcal{L}_{e,j}}
\alpha_{e,j,\ell}
\tilde{\mathbf{r}}_{e,j,\ell}^{\mathrm{path}}.
\label{eq:channel_representation}
\end{equation}
The per-channel position head predicts a conditional density over the BS-centered UE position, $p_{\Theta}^{\mathrm{ch}}(\mathbf{p}\,|\,\mathbf{r}_{e,j}^{\mathrm{ch}})$, where $\mathbf{p}\in\mathbb{R}^{3}$ denotes a UE position. The density $p_{\Theta}^{\mathrm{ch}}(\cdot)$ is also parameterized as a GMM, whose components are predicted from $\mathbf{r}_{e,j}^{\mathrm{ch}}$ using an MLP head with one hidden layer. During training, the predicted mixture density is evaluated at the observed query UE position $\hat{\mathbf{p}}_{e,j}$. The per-path position head models the ambiguity of individual propagation paths, while the per-channel position head combines multiple environment-conditioned path representations to provide the final channel-level position distribution.

\subsection{Training Objective}
\label{subsec:c2w_training_method}

\begin{figure}[t]
	\begin{center}
		{\includegraphics[width=0.87\columnwidth,keepaspectratio]
			{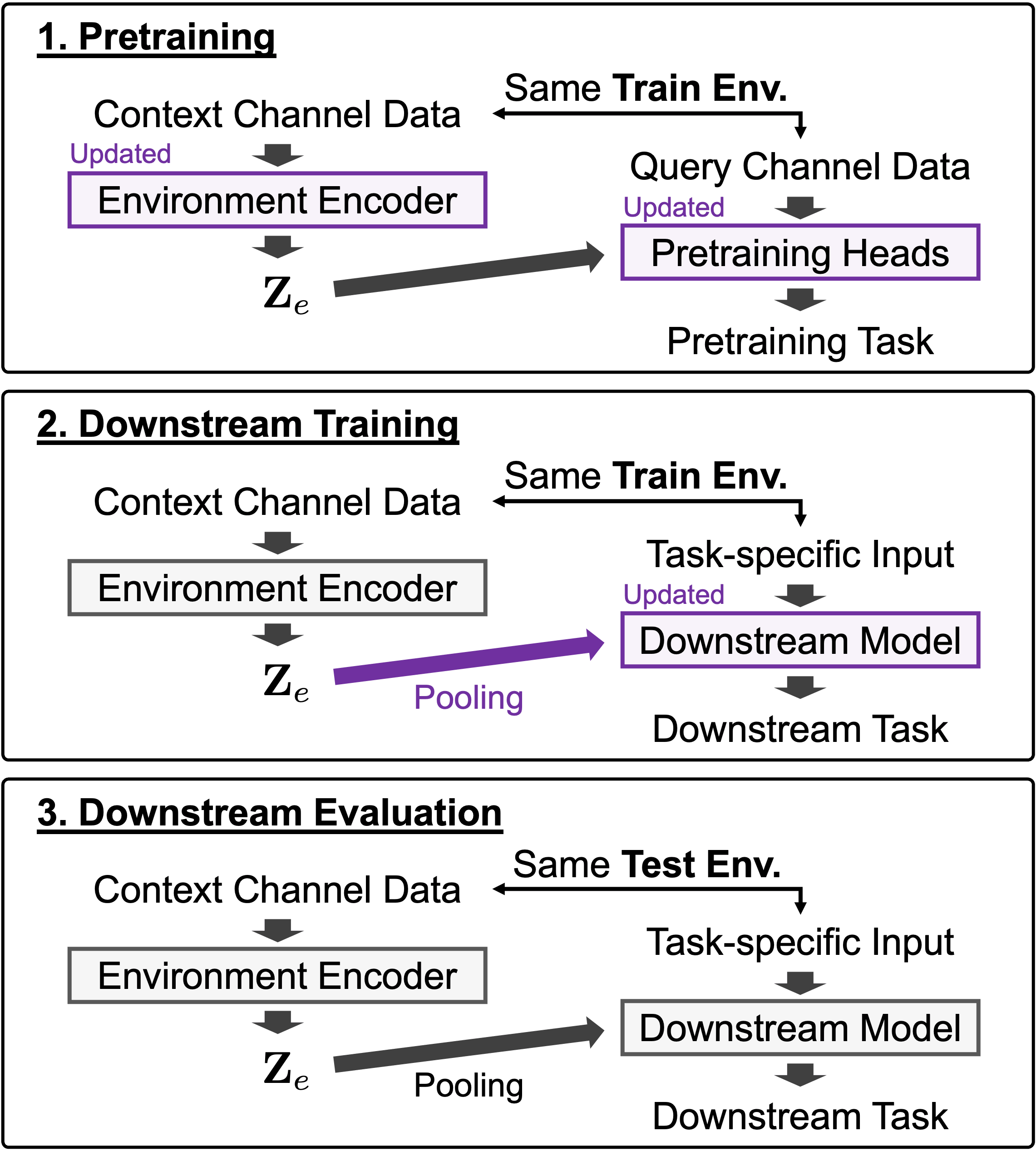}%
			\caption{Pretraining and downstream protocol. The encoder is pretrained with context-query channel sets and then frozen for task-specific downstream learning.}
			\label{fig:protocol}
		}
	\end{center}
	\vspace{-10pt}
\end{figure}

Pretraining is performed using the three prediction tasks described in Section~\ref{subsec:c2w_query_prediction}. Let $\mathcal{E}_{\mathrm{tr}},\mathcal{E}_{\mathrm{te}}\subset\{1,\cdots,E\}$ denote the training and test environment sets, respectively. For each sampled environment $e\in\mathcal{E}_{\mathrm{tr}}$, we construct disjoint context and query channel-index sets satisfying $\mathcal{C}_{e}\cap\mathcal{Q}_{e}=\emptyset$. Let $N_{\mathrm{path}}^{\mathrm{qry}}=\sum_{j\in\mathcal{Q}_{e}}|\mathcal{L}_{e,j}|$ denote the number of query path tokens, where $|\mathcal{L}_{e,j}|$ is the cardinality of the estimated path-index set for query channel $j$. The per-path position loss, relative path-gain loss, and per-channel position loss are defined as
\begin{equation}
\begin{aligned}
    \mathcal{L}_{\mathrm{pos}}^{\mathrm{path}}
    &=
    -
    \frac{1}{N_{\mathrm{path}}^{\mathrm{qry}}}
    \sum_{j\in\mathcal{Q}_{e}}
    \sum_{\ell\in\mathcal{L}_{e,j}}
    \log
    p_{\Theta}^{\mathrm{path}}
    \big(
    \hat{\mathbf{p}}_{e,j}
    \,|\,
    \mathbf{r}_{e,j,\ell}^{\mathrm{path}}
    \big),\\
    \mathcal{L}_{\mathrm{gain}}
    &=
    \frac{1}{N_{\mathrm{path}}^{\mathrm{qry}}}
    \sum_{j\in\mathcal{Q}_{e}}
    \sum_{\ell\in\mathcal{L}_{e,j}}
    \big(
    \hat{\bar{g}}_{e,j,\ell}^{\mathrm{pred}}
    -
    \hat{\bar{g}}_{e,j,\ell}
    \big)^2,\\
    \mathcal{L}_{\mathrm{pos}}^{\mathrm{ch}}
    &=
    -
    \frac{1}{|\mathcal{Q}_{e}|}
    \sum_{j\in\mathcal{Q}_{e}}
    \log
    p_{\Theta}^{\mathrm{ch}}
    \big(
    \hat{\mathbf{p}}_{e,j}
    \,|\,
    \mathbf{r}_{e,j}^{\mathrm{ch}}
    \big).
\end{aligned}
\label{eq:pretraining_losses}
\end{equation}
The total loss for one environment-level task is
\begin{equation}
    \mathcal{L}_{e}
    =
    \mathcal{L}_{\mathrm{pos}}^{\mathrm{path}}
    +
    \lambda_{\mathrm{g}}
    \mathcal{L}_{\mathrm{gain}}
    +
    \lambda_{\mathrm{ch}}
    \mathcal{L}_{\mathrm{pos}}^{\mathrm{ch}},
    \label{eq:total_training_loss}
\end{equation}
where $\lambda_{\mathrm{g}}>0$ and $\lambda_{\mathrm{ch}}>0$ are loss weights for relative path-gain estimation and channel-level position estimation, respectively. The training objective is
\begin{equation}
    \min_{\Theta}
    \;
    \mathbb{E}_{e\sim\mathcal{E}_{\mathrm{tr}}}
    \mathbb{E}_{\mathcal{C}_{e},\mathcal{Q}_{e}}
    \left[
    \mathcal{L}_{e}
    \right],
    \label{eq:training_objective}
\end{equation}
where the outer expectation is over sampled training environments and the inner expectation is over the sampling of context and query channel-index sets within each environment. The overall training and evaluation protocol is summarized in Fig.~\ref{fig:protocol}. During pretraining, the environment encoder and the pretraining heads are jointly optimized. For each downstream task, the pretrained environment encoder is frozen and used to provide the wireless world embedding $\mathbf{Z}_{e}$, while only the task-specific downstream model is updated. During downstream evaluation, both the encoder and the downstream model are fixed, and generalization performance is evaluated using data from fully unseen test environments $\mathcal{E}_{\mathrm{te}}$.

\section{Downstream Tasks}
\label{sec:downstream_tasks}

After pretraining, the frozen {\sffamily\scshape Channel2World} encoder produces a wireless world embedding $\mathbf{Z}_{e}\in\mathbb{R}^{K_{\mathrm{z}}\times D_{\mathrm{m}}}$ for environment $e$. For a downstream task, let $\mathbf{o}_{e,i}$ denote the task-specific observation associated with UE channel $i$, and let $\hat{\mathbf{y}}_{e,i}$ denote the predicted target. A latent-conditioned downstream predictor is written as
\begin{equation}
    \hat{\mathbf{y}}_{e,i}
    =
    f_{\Psi}
    \big(
    \mathbf{o}_{e,i},
    \mathbf{Z}_{e}
    \big),
    \label{eq:general_downstream_model}
\end{equation}
where $\Psi$ denotes the task-specific trainable parameters. We consider three general downstream tasks: UE position estimation from a single spatial channel matrix, beam-domain CSI reconstruction from sparse beam measurements, and RF-observable geometry reconstruction from NLoS path directions. These tasks evaluate whether the wireless world embedding provides reusable environment conditioning across localization, communication, and geometric inference.

\subsection{Environment-Aware Channel Matrix-to-Position Estimation}
\label{subsec:downstream_channel_position}

The first downstream task evaluates whether the wireless world embedding can assist UE position estimation from a single spatial channel matrix. This setting differs from the pretraining task, where path-level delay and AoA are explicitly available. For UE channel $i$ in environment $e$, the downstream observation is defined as the input spatial channel matrix, $\mathbf{H}_{e,i}\in\mathbb{C}^{N_x\times N_y}$. The target is the BS-centered UE position $\mathbf{p}_{e,i}\in\mathbb{R}^{3}$. The latent-conditioned position estimator is
\begin{equation}
    \mathbf{p}_{e,i}^{\mathrm{pred}}
    =
    f_{\Psi}^{\mathrm{pos}}
    \big(
    \mathbf{H}_{e,i},
    \mathbf{Z}_{e}
    \big),
    \label{eq:downstream_position_estimator}
\end{equation}
where $f_{\Psi}^{\mathrm{pos}}(\cdot)$ denotes a trainable estimator. The complex channel matrix $\mathbf{H}_{e,i}$ is converted into a real-valued feature vector by vectorizing its entries and concatenating their real and imaginary parts. This channel feature is concatenated with an attention-pooled representation of $\mathbf{Z}_{e}$ and passed to an MLP prediction head, which outputs $\mathbf{p}_{e,i}^{\mathrm{pred}}$. The training loss is defined as the mean squared position error,
%This channel feature is fused with an environment-context feature derived from $\mathbf{Z}_{e}$ using attention pooling, and the resulting feature is passed to an MLP prediction head that outputs $\mathbf{p}_{e,i}^{\mathrm{pred}}$.
\begin{equation}
    \mathcal{L}_{\mathrm{pos}}^{\mathrm{ds}}
    =
    \mathbb{E}
    \big[
    \|
    \mathbf{p}_{e,i}^{\mathrm{pred}}
    -
    \mathbf{p}_{e,i}
    \|_{2}^{2}
    \big].
    \label{eq:downstream_position_loss}
\end{equation}
This task evaluates whether $\mathbf{Z}_{e}$ provides site-specific environmental context for localization beyond the information contained in a direct channel-matrix observation.

\subsection{Environment-Aware Beam-Domain CSI Reconstruction}
\label{subsec:downstream_beam_reconstruction}

The second downstream task addresses beam-domain downlink CSI reconstruction using partial beam-gain feedback from the UE. In practical beam management, the BS may observe only a subset of beam measurements due to pilot and feedback overhead. The objective is to reconstruct the full 2D DFT beam-domain received-power map from sparse, noisy beam observations. Such reconstruction is useful for beam management in multiuser MIMO systems that require interference-aware beam selection and channel-state prediction~\cite{moon2026fr3beam}.

For this downstream task, all environments are evaluated under the same BS array configuration and the same fixed 2D DFT beamforming vector set $\mathcal{B}=\{\mathbf{f}_{m}\}_{m=1}^{M}$, where $\mathbf{f}_{m}\in\mathbb{C}^{N_xN_y}$ and $M=N_xN_y$. For UE channel $i$ in environment $e$, let $\boldsymbol{\gamma}_{e,i}\in\mathbb{R}^{M}$ denote the full downlink beam-domain received-power vector in dBm, whose $m$-th entry is the received power at the UE when the BS uses beam $\mathbf{f}_{m}$. The full vector $\boldsymbol{\gamma}_{e,i}$ is the reconstruction target. Let $\Omega\subset\{1,\cdots,M\}$ denote the fixed set of measured codebook beam indices. The sparse beam observation $\boldsymbol{\gamma}_{e,i}^{\Omega}\in\mathbb{R}^{M}$ is defined as
\begin{equation}
    [\boldsymbol{\gamma}_{e,i}^{\Omega}]_{m}
    =
    \begin{cases}
    \tilde{\gamma}_{e,i,m}, & m\in\Omega,\\
    0, & m\notin\Omega,
    \end{cases}
    \label{eq:sparse_beam_observation}
\end{equation}
where $\tilde{\gamma}_{e,i,m}$ denotes the noisy dBm-scale beam-power measurement for beam $m$, obtained under the assumption of an isotropic single-antenna UE. The environment-aware beam-domain CSI reconstruction model is
\begin{equation}
    \hat{\boldsymbol{\gamma}}_{e,i}
    =
    f_{\Psi}^{\mathrm{beam}}
    \big(
    \boldsymbol{\gamma}_{e,i}^{\Omega},
    \mathbf{Z}_{e}
    \big),
    \label{eq:beam_reconstruction_model}
\end{equation}
where $f_{\Psi}^{\mathrm{beam}}(\cdot)$ denotes a convolutional neural network (CNN)-based latent-conditioned CSI reconstructor and $\hat{\boldsymbol{\gamma}}_{e,i}\in\mathbb{R}^{M}$ is the reconstructed full beam-domain received-power vector in dBm. The sparse dBm-scale beam vector is first reshaped onto the 2D DFT beam grid and processed by convolutional layers to extract beam-domain features. In parallel, the wireless world embedding $\mathbf{Z}_{e}$ is converted into an environment-context feature through attention pooling over latent tokens. The beam-domain and environment-context features are then concatenated and passed to an MLP head, which predicts the full dBm-scale beam-domain output. The supervised objective is
\begin{equation}
    \mathcal{L}_{\mathrm{beam}}^{\mathrm{ds}}
    =
    \mathbb{E}
    \big[
    \|
    \hat{\boldsymbol{\gamma}}_{e,i}
    -
    \boldsymbol{\gamma}_{e,i}
    \|_{2}^{2}
    \big].
    \label{eq:beam_reconstruction_loss}
\end{equation}
This task evaluates whether the wireless world embedding can also support site-specific communication tasks that are not explicitly formulated as geometric inference problems.

\subsection{Environment Reconstruction via Interaction Point Estimation}
\label{subsec:downstream_environment_reconstruction}

The third downstream task assesses whether the wireless world embedding, beyond serving as a conditioning variable for communication tasks, contains recoverable information about the RF-observable geometry of the environment. This task reconstructs the first BS-side interaction points associated with NLoS paths. Given the BS-centered AoA of a path, the model estimates the range to the first interaction point along the corresponding ray direction. The resulting point set provides a BS-centered partial reconstruction of the propagation-relevant environment geometry.

\begin{figure}[t]
	\begin{center}
		{\includegraphics[width=0.95\columnwidth,keepaspectratio]
			{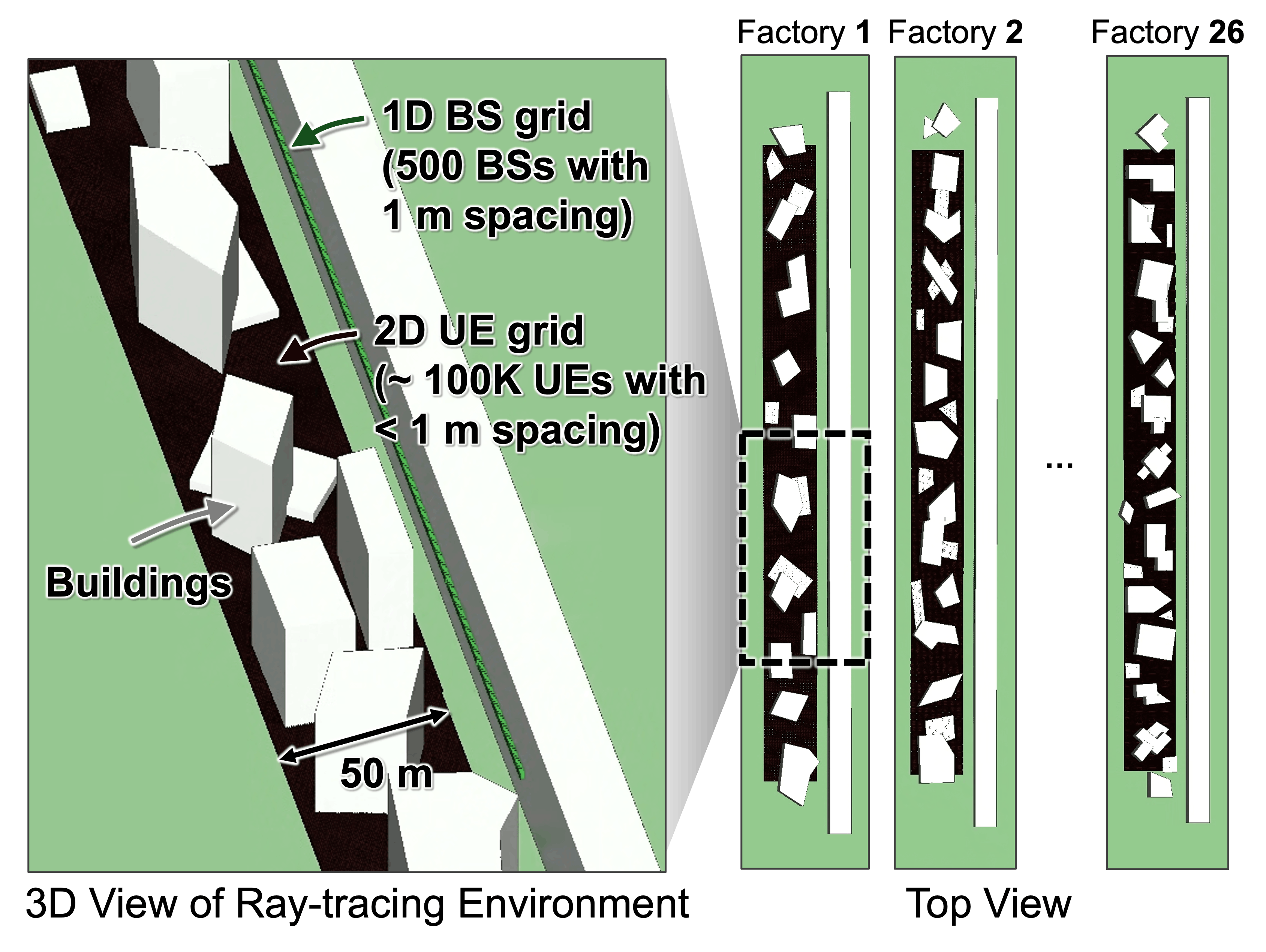}%
			\caption{Ray-tracing factory dataset. A one-dimensional BS grid and a dense two-dimensional UE grid generate BS-centered propagation environments for each layout.}
			\label{fig:rtenv}
		}
	\end{center}
	\vspace{-10pt}
\end{figure}

For an NLoS path $\ell$ of UE channel $i$ in environment $e$, let $\mathbf{s}_{e,i,\ell}^{(1)}\in\mathbb{R}^{3}$ denote the first BS-side interaction point in the BS-centered coordinate system, and let $r_{e,i,\ell}^{(1)}=\|\mathbf{s}_{e,i,\ell}^{(1)}\|$ denote its range from the BS. The path-level observation is the BS-centered AoA feature
\begin{equation}
    \boldsymbol{\eta}_{e,i,\ell}
    =
    [
    \sin\hat{\theta}_{e,i,\ell},
    \cos\hat{\theta}_{e,i,\ell},
    \sin\hat{\phi}_{e,i,\ell},
    \cos\hat{\phi}_{e,i,\ell}
    ]^{\mathrm{T}} .
    \label{eq:downstream_aoa_feature}
\end{equation}
The latent-conditioned range predictor $p_{\Psi}^{\mathrm{rec}}(\cdot)$ is modeled as a GMM-based one-dimensional probabilistic estimator,
\begin{equation}
    p_{\Psi}^{\mathrm{rec}}
    \big(
    r
    \,|\,
    \boldsymbol{\eta}_{e,i,\ell},
    \mathbf{Z}_{e}
    \big),
    \label{eq:interaction_range_distribution}
\end{equation}
and is trained using the negative log-likelihood
\begin{equation}
    \mathcal{L}_{\mathrm{rec}}^{\mathrm{ds}}
    =
    -
    \mathbb{E}
    \big[
    \log
    p_{\Psi}^{\mathrm{rec}}
    \big(
    r_{e,i,\ell}^{(1)}
    \,|\,
    \boldsymbol{\eta}_{e,i,\ell},
    \mathbf{Z}_{e}
    \big)
    \big],
    \label{eq:environment_reconstruction_loss}
\end{equation}
where the expectation is taken over the NLoS path samples from training environments. Given the predicted range distribution, the point estimate is obtained by maximum likelihood:
\begin{equation}
    \hat{r}_{e,i,\ell}^{(1)}
    =
    \operatorname*{argmax}_{r}
    \,
    p_{\Psi}^{\mathrm{rec}}
    \big(
    r
    \,|\,
    \boldsymbol{\eta}_{e,i,\ell},
    \mathbf{Z}_{e}
    \big).
    \label{eq:interaction_range_ml_estimate}
\end{equation}
The first interaction point is then reconstructed along the AoA-defined ray. Let $\mathbf{u}_{e,i,\ell}^{\mathrm{AoA}}\in\mathbb{R}^{3}$ denote the unit direction vector associated with the BS-centered AoA convention,
\begin{equation}
    \mathbf{u}_{e,i,\ell}^{\mathrm{AoA}}
    =
    [
    \sin\hat{\theta}_{e,i,\ell}\cos\hat{\phi}_{e,i,\ell},
    \sin\hat{\theta}_{e,i,\ell}\sin\hat{\phi}_{e,i,\ell},
    \cos\hat{\theta}_{e,i,\ell}
    ]^{\mathrm{T}} .
    \label{eq:aoa_unit_direction}
\end{equation}
The reconstructed interaction point is
\begin{equation}
    \hat{\mathbf{s}}_{e,i,\ell}^{(1)}
    =
    \hat{r}_{e,i,\ell}^{(1)}
    \mathbf{u}_{e,i,\ell}^{\mathrm{AoA}} .
    \label{eq:reconstructed_interaction_point}
\end{equation}
Applying this procedure to multiple NLoS paths yields a sparse reconstruction of the RF-observable environment geometry.

\begin{table}[t]
\centering
\caption{Key simulation and model parameters.}
\label{tab:sim_parameters}
\footnotesize
\begin{tabular}{p{0.50\columnwidth}p{0.39\columnwidth}}
\hline
Parameter & Value \\
\hline
Factory layouts & $26$ randomized layouts \\
Factory-level train/validation/test split & $22$/$2$/$2$ \\
BS per layout & $500$ positions \\
Total BS-centered environments & $26000$ \\
BS height and wall offset & $15$~m and $0.2$~m \\
UE height & $1.5$~m \\
UE grid spacing & Randomized in $[0.6,0.7]$~m \\
Carrier frequency & $7.0$~GHz \\
Maximum paths & $12$ per channel \\
Maximum path interactions & $3$ \\
Environment latent tokens & $K_{\mathrm{z}}=16$\\
Token dimension & $D_{\mathrm{m}}=128$ \\
Context and query channels & $128$ and $32$ \\
Attention heads & $4$ \\
Path self-attention layers & $4$ \\
Encoder cross-attention layers & $6$ \\
Encoder self-attention layers & $6$ \\
Query cross-attention layers & $8$ \\
Query path self-attention layers & $2$ \\
GMM components & $8$ \\
Minimum GMM std. & $\sigma_{\min}=0.5$~m \\
Pretraining batch size & $32$ \\
Loss weights & $\lambda_{\mathrm{g}}=0.1$ and $\lambda_{\mathrm{ch}}=0.5$ \\
Error-aware position observation error & $\sigma_{\mathrm{pos}}=0.1$~m \\
Error-aware delay error & $\sigma_{\tau}=1$~ns \\
Error-aware angle errors & $\sigma_{\theta}=\sigma_{\phi}=1^{\circ}$ \\
UE-side gain fluctuation & $\sigma_{\xi}=3$~dB\\
\hline
\end{tabular}
\vspace{-10pt}
\end{table}

\section{Simulation Results}
\label{sec:simulation_results}

This section evaluates {\sffamily\scshape Channel2World} using ray-tracing channel data generated from $26000$ BS-centered factory environments.
%The evaluation follows the protocol in Fig.~\ref{fig:protocol}. The {\sffamily\scshape Channel2World} encoder is first pretrained using path-domain channel-position observations and then frozen. The frozen encoder provides the wireless world embedding $\mathbf{Z}_{e}$ for downstream tasks, while each downstream model only trains its own task-specific parameters. Downstream performance is evaluated on test environments that are not used for either encoder pretraining or downstream-model training.
Table~\ref{tab:sim_parameters} summarizes the main dataset, model, and training parameters. The pretrained model uses $K_{\mathrm{z}}=16$ environment latent tokens with token dimension $D_{\mathrm{m}}=128$. Each pretraining sample contains $128$ context channels, jointly processed as input, and $32$ query channels from the same BS-centered environment. The environment encoder uses four per-channel path self-attention layers, followed by six alternating bidirectional cross-attention and self-attention layers for environment-token aggregation. The estimation heads use eight bidirectional cross-attention layers and two post-processing path self-attention layers. All attention blocks follow a Transformer structure with residual connections, four attention heads, a feed-forward network (FFN) expansion ratio of 4, and no dropout~\cite{vaswani2017attention}. With these settings, {\sffamily\scshape Channel2World} contains approximately $9.2$ million trainable parameters.

We consider two pretraining conditions. In the \emph{error-aware} condition, the decomposition-equivalent error model in~\eqref{eq:path_estimation_error} is applied to the path parameters, and the UE position error model in~\eqref{eq:position_error} is applied to the observed positions. The delay-error standard deviation is $\sigma_{\tau}=1$~ns, and the AoA-error standard deviations are $\sigma_{\theta}=\sigma_{\phi}=1^{\circ}$. The observed UE positions are generated with $\sigma_{\mathrm{pos}}=0.1$~m. The UE-side path-gain fluctuation in~\eqref{eq:practical_path_power} is set to $\sigma_{\xi}=3$~dB. In the \emph{error-free} condition, the delay, angle, and position perturbations are removed, while the same $\sigma_{\xi}=3$~dB gain fluctuation is applied.

\begin{figure}[t]
	\begin{center}
		{\includegraphics[width=0.8\columnwidth,keepaspectratio]
			{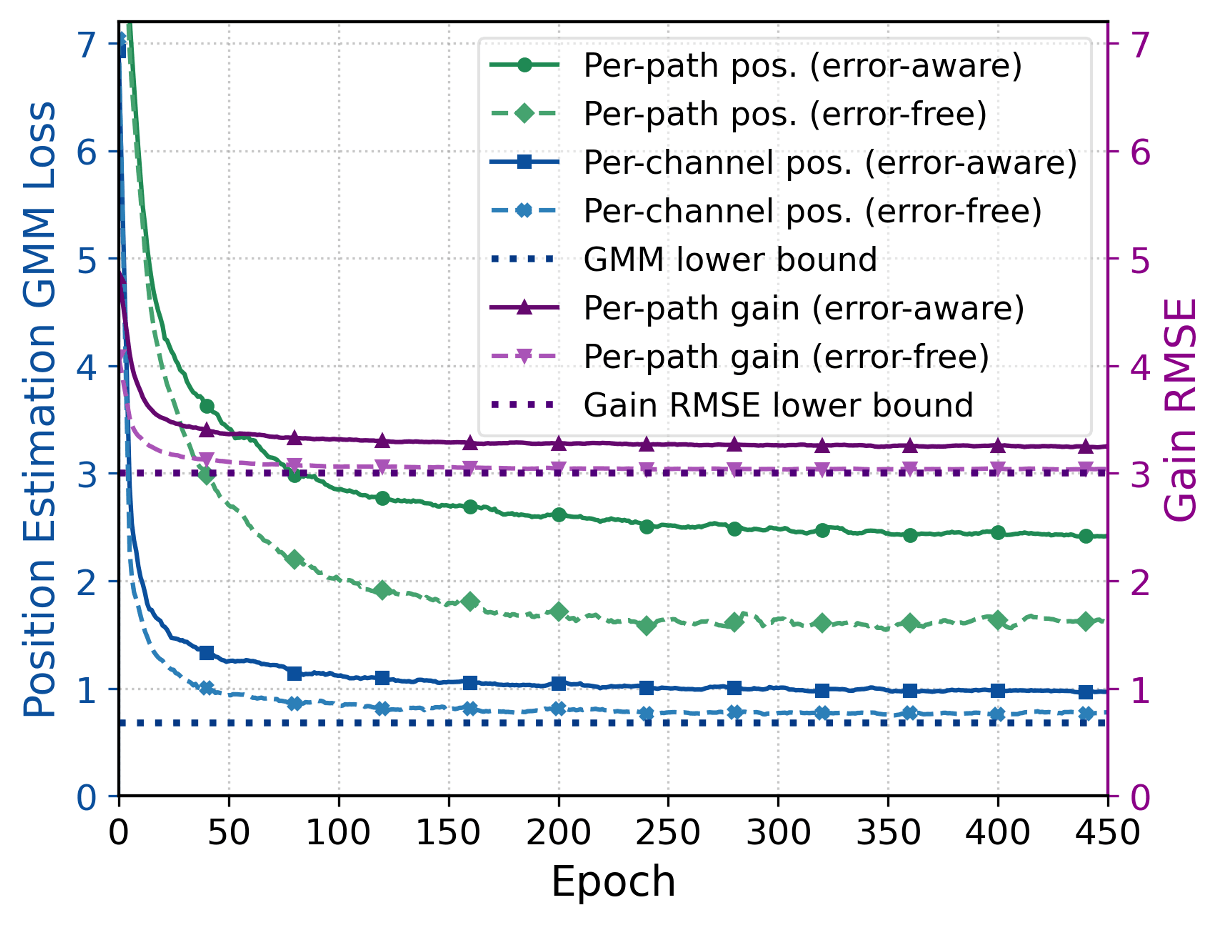}%
			\caption{Pretraining validation curves for position GMM loss and relative gain RMSE under error-aware and error-free inputs.}
			\label{fig:res03}
		}
	\end{center}
	\vspace{-10pt}
\end{figure}

\begin{figure}[t]
	\begin{center}
		{\includegraphics[width=1.0\columnwidth,keepaspectratio]
			{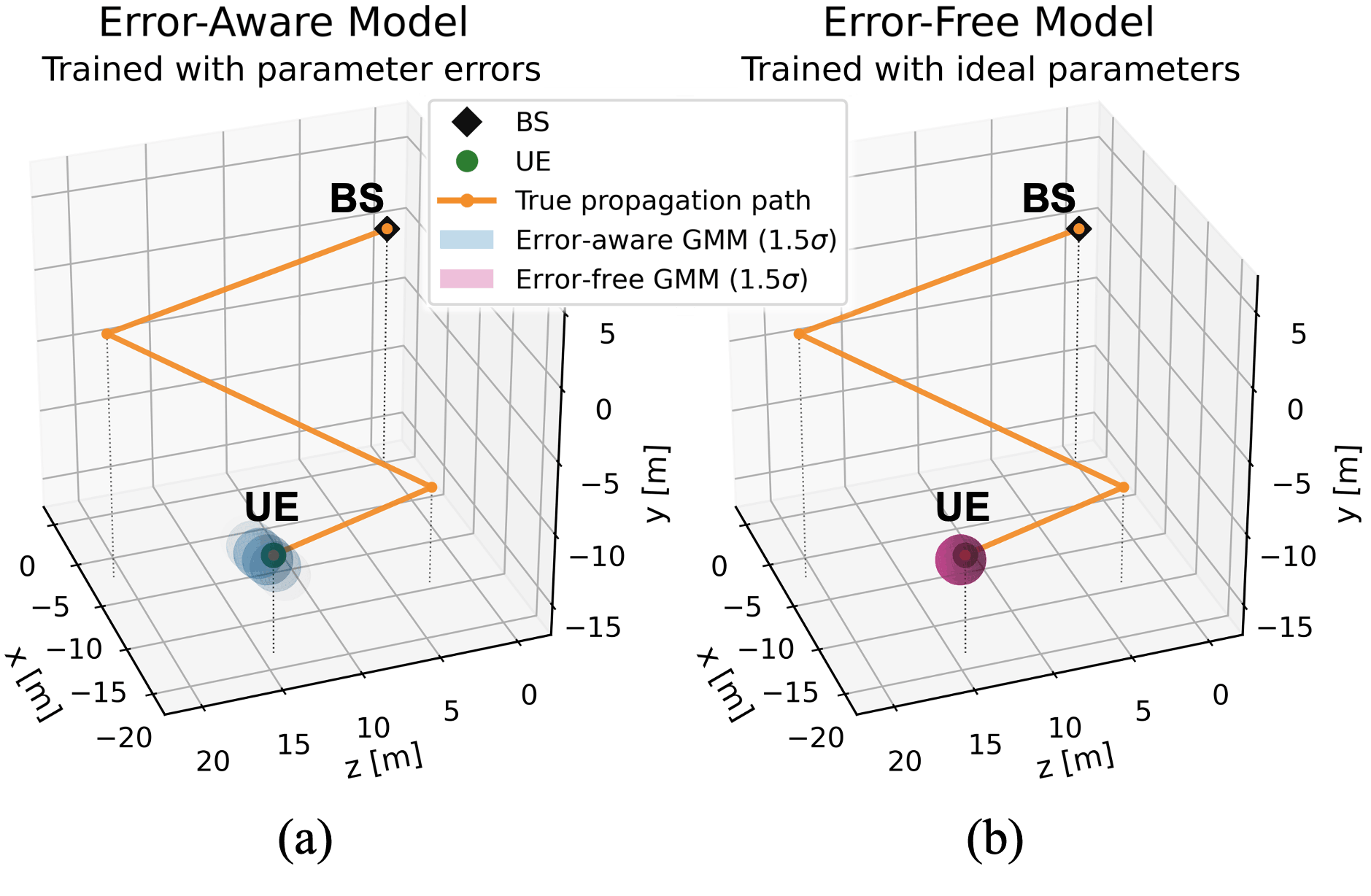}%
			\caption{Per-path UE-position GMMs for a query path in a test environment.}
			\label{fig:res06}
		}
	\end{center}
	\vspace{-10pt}
\end{figure}

\subsection{Dataset Generation}
\label{subsec:dataset_generation}

The dataset is generated using Remcom Wireless InSite ray-tracing software. Fig.~\ref{fig:rtenv} shows one representative factory layout. We generate $26$ base layouts with randomized building geometry. For each layout, $500$ BS positions are placed along a large wall with $1$~m spacing. Each BS is located at a height of $15$~m with a $0.2$~m offset from the wall. UE positions are sampled over the service region using a dense 2D grid, whose spacing is randomized between $0.6$~m and $0.7$~m to reduce grid-induced learning artifacts.

For a given factory layout and BS position, the set of valid UE channels defines a single BS-centered propagation environment. Diffraction is disabled in the ray-tracing simulation, and reflection paths with up to three reflections are considered. We retain up to the $12$ strongest paths per channel and remove channels with no valid propagation path. After filtering, each BS-centered environment contains $833$ to $9024$ valid UE channel-position pairs, with an average of $4999$ pairs. Horizontal mirror augmentation is then applied by reversing the $x_e$-axis and transforming the corresponding UE positions and path parameters. The final dataset contains $26\times500\times2=26000$ unique BS-centered environments. The train/validation/test environment split is performed at the factory-layout level rather than at the BS level. This prevents neighboring BS-centered scenes, mirrored counterparts, and UE samples from the same layout from appearing across different splits. Hence, the downstream results evaluate generalization to unseen building realizations, rather than interpolation among BS locations from previously observed layouts.

\subsection{Channel2World Pretraining}
\label{subsec:sim_pretraining}

Fig.~\ref{fig:res03} shows the validation performance of {\sffamily\scshape Channel2World} over training epochs for the three pretraining objectives. The two position curves represent the GMM negative log-likelihoods of the per-path and per-channel position heads, while the gain curve represents the root mean squared error (RMSE) of the relative path-gain estimate. The dotted GMM lower bound is obtained from the minimum component standard deviation $\sigma_{\min}=0.5$~m. For a 3D spherical Gaussian component centered at the target, the minimum negative log-likelihood is $\frac{3}{2}\log(2\pi\sigma_{\min}^{2})$. The gain-RMSE lower bound is $\sigma_{\xi}=3$~dB, since the relative gain target includes an independent UE-side fluctuation that cannot be inferred from the channel-position context.

Fig.~\ref{fig:res06} visualizes the per-path position GMM produced by the pretraining head for a test environment. The query token contains only the calibrated propagation-delay estimate and BS-centered AoA of a single path. For the illustrated NLoS path, these two quantities do not uniquely determine the UE position, since the inverse mapping depends on the reflector geometry of the environment. Nevertheless, both the error-aware and error-free models place GMM components near the target UE position, even in an unseen test environment. This result indicates that $\mathbf{Z}_{e}$, inferred from multiple context channel tokens, provides the environmental information needed to interpret the delay and AoA features through the surrounding reflector structure. The error-aware model produces a broader distribution, consistent with the injected perturbations, while the error-free model yields a sharper estimate under ideal path inputs.

\begin{figure}[t]
	\begin{center}
		{\includegraphics[width=0.8\columnwidth,keepaspectratio]
			{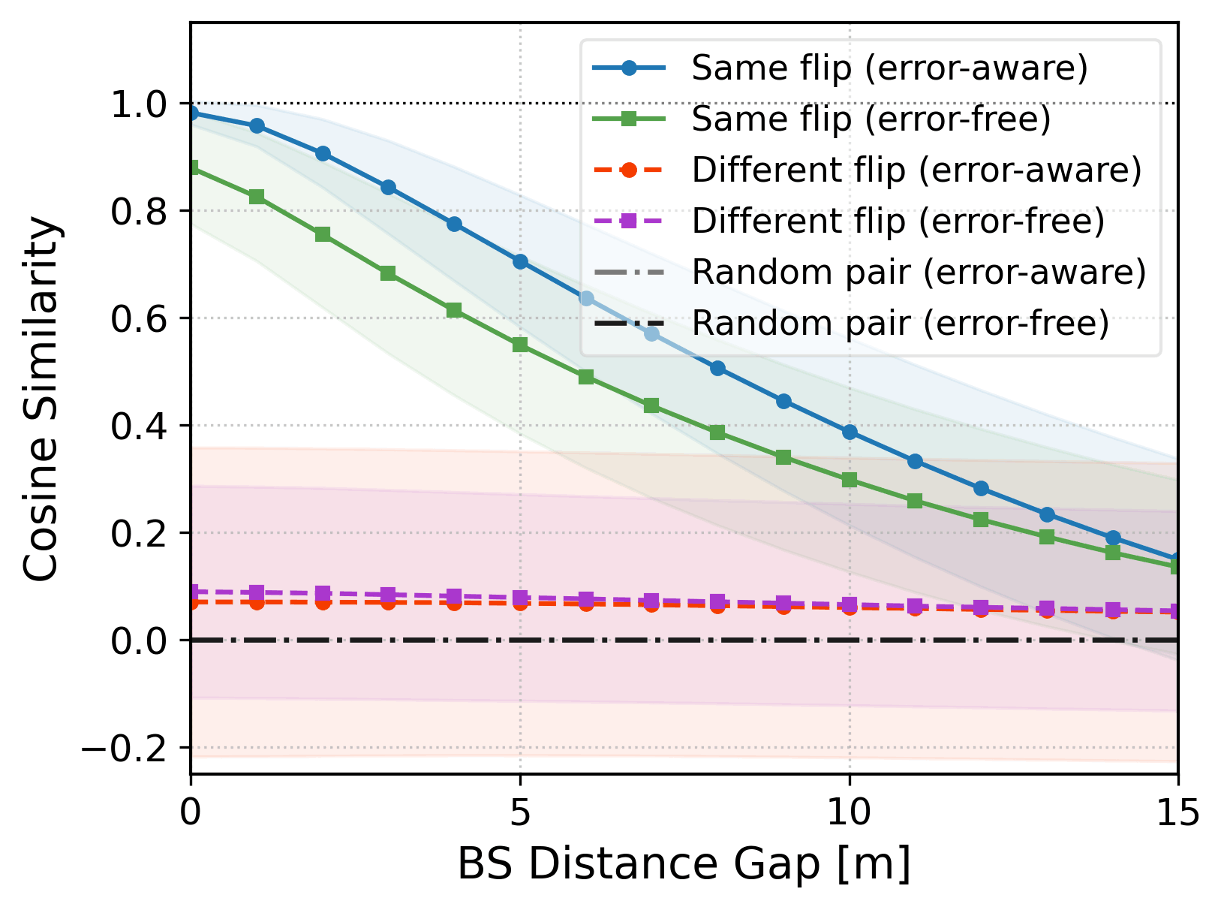}%
			\caption{Cosine similarity between wireless world embeddings obtained from nearby BSs in the same factory layout.}
			\label{fig:res02}
		}
	\end{center}
	\vspace{-10pt}
\end{figure}

\subsection{Latent Continuity Across Environments}
\label{subsec:sim_latent_analysis}

Fig.~\ref{fig:res02} analyzes the latent geometry of the learned wireless world embeddings. For this analysis, each embedding is inferred from $1024$ randomly sampled context channels from the corresponding BS-centered environment. We subtract the empirical mean of all $26000$ environment embeddings from each embedding and compute cosine similarity between the centered embeddings. The random-pair reference lines indicate the mean similarity of randomly paired environments under the same centering and normalization procedure. Their values are close to zero, showing that the observed positive similarity of nearby same-flip environments is not caused by a global embedding offset.

\begin{figure}[t]
	\begin{center}
		{\includegraphics[width=1.0\columnwidth,keepaspectratio]
			{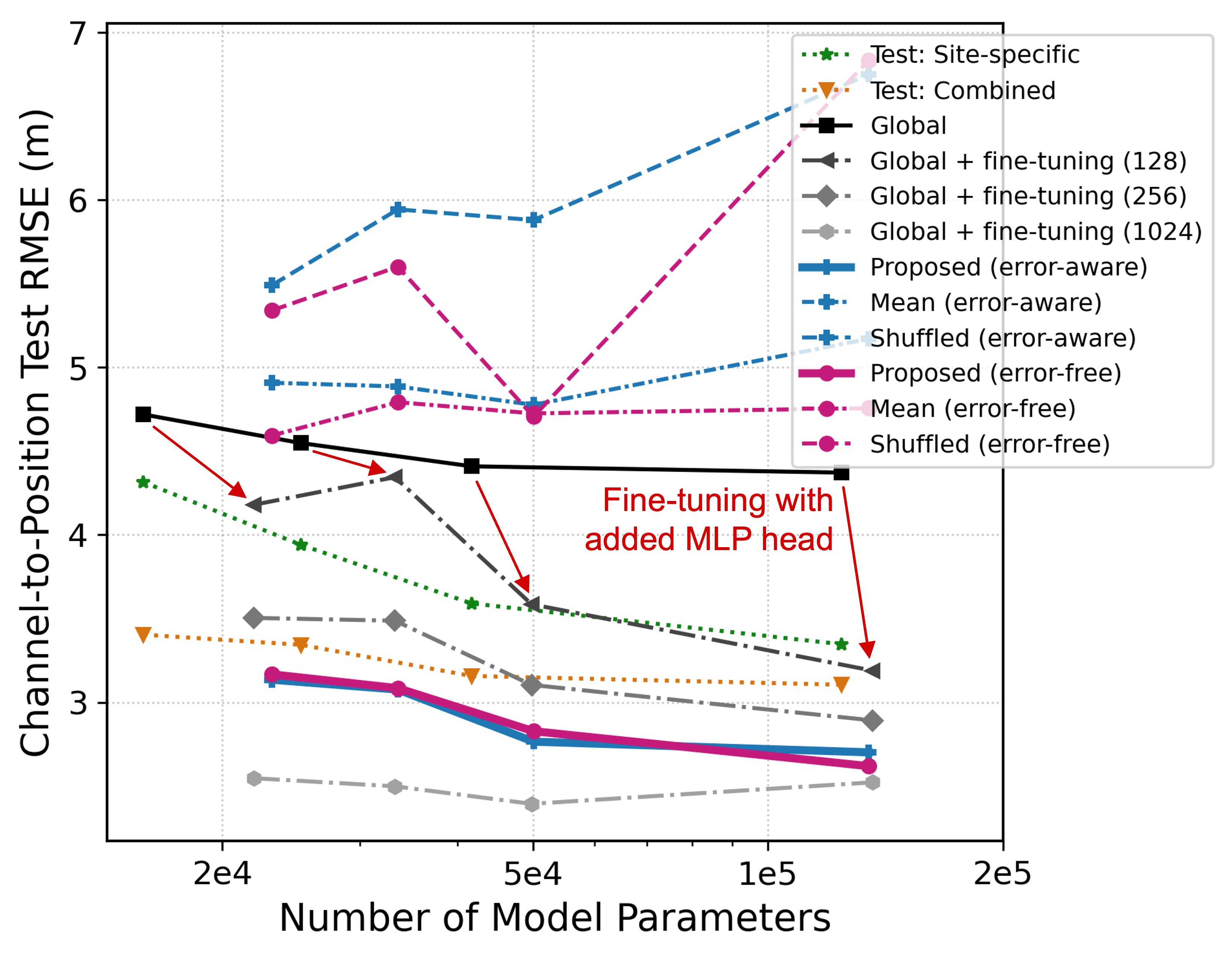}%
			\caption{Channel matrix-to-position estimation in test environments. Position RMSE is shown versus the number of downstream-model parameters.}
			\label{fig:res04}
		}
	\end{center}
	\vspace{-10pt}
\end{figure}

The same-flip curves compare BS-centered environments from the same factory layout and mirror-augmentation sign. Their similarity decreases with BS separation, as adjacent BSs observe partially overlapping reflector and blockage structures, while distant BSs induce more distinct BS-centered propagation relationships. At zero distance gap, the embeddings correspond to the same environment but are inferred from different random context-channel subsets, which explains the small deviation from unit similarity. The different-flip curves remain close to the random-pair reference. Although a mirror-flipped environment may share coarse layout complexity with the original environment, its BS-centered geometry and AoA-to-reflector correspondence are distinct. The embeddings produced by the error-aware encoder generally show higher same-flip similarity than those produced by the error-free encoder, suggesting that error-aware pretraining learns a smoother embedding space under local perturbations. In contrast, the error-free encoder preserves finer site-specific differences, yielding sharper separation between nearby BS-centered environments.

\begin{figure}[t]
	\begin{center}
		{\includegraphics[width=1.0\columnwidth,keepaspectratio]
			{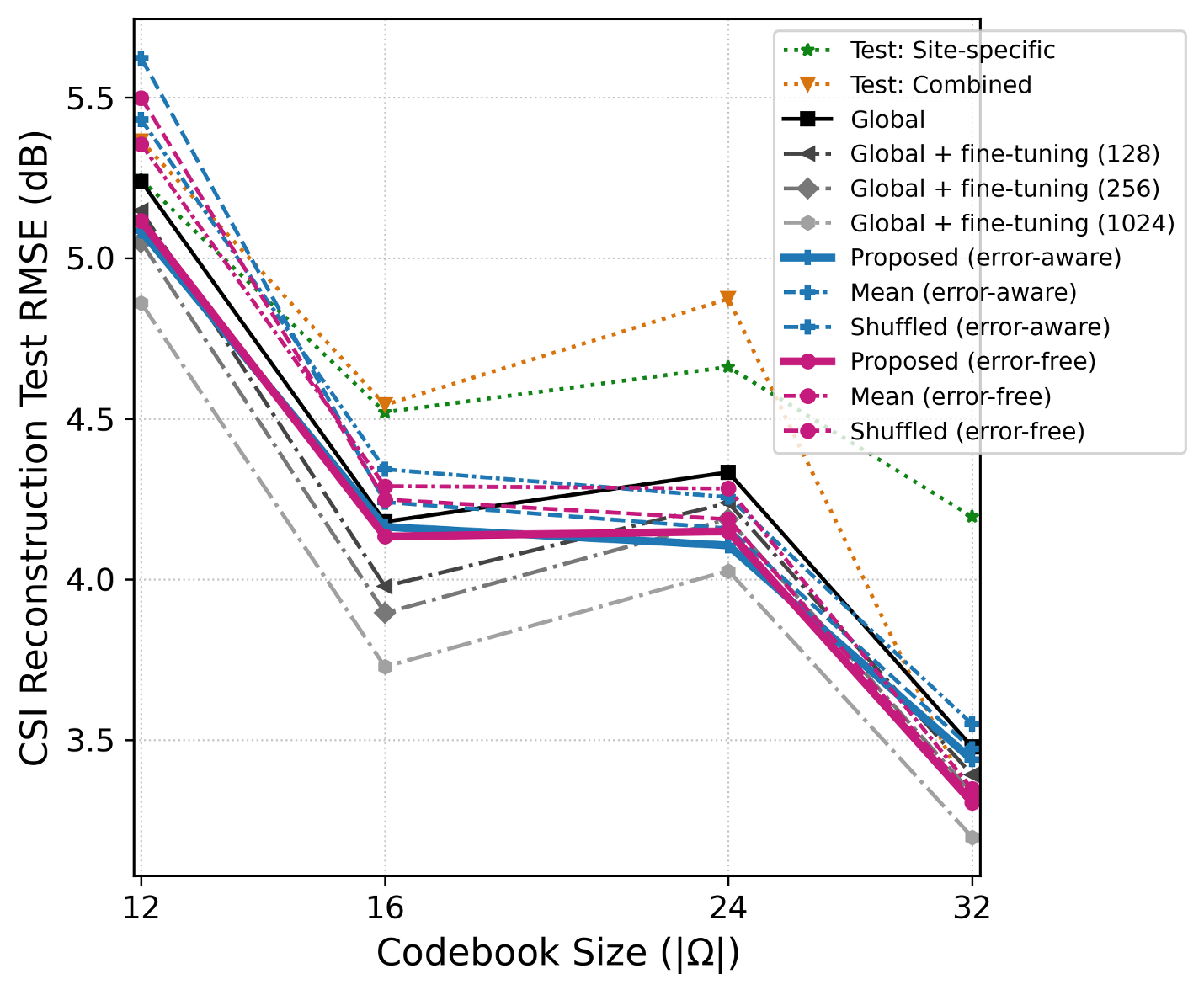}%
			\caption{Beam-domain CSI reconstruction RMSE from sparse codebook measurements in test environments.}
			\label{fig:res05}
		}
	\end{center}
	\vspace{-10pt}
\end{figure}

\begin{figure*}[t]
    \centering
    \includegraphics[width=1.96\columnwidth]{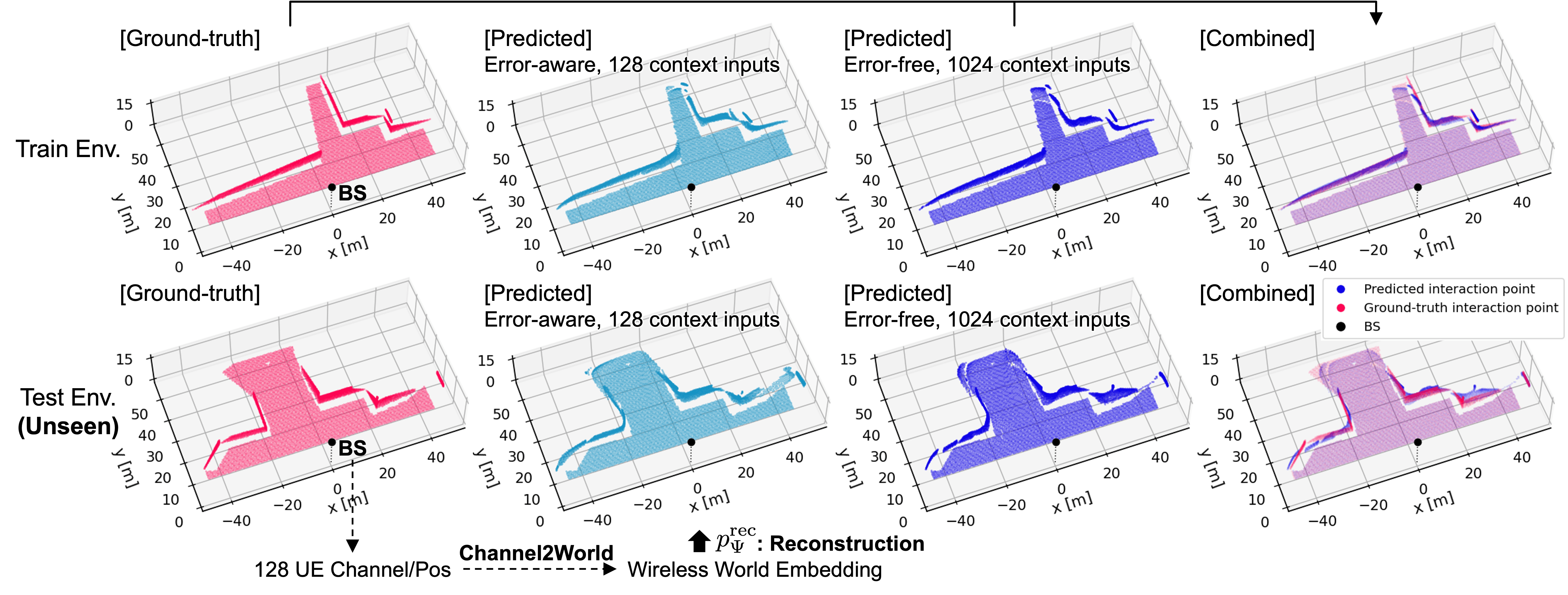}
    \caption{RF-observable geometry reconstruction from NLoS path directions. Predicted interaction points are shown using embeddings inferred by the error-aware encoder with $128$ context channels and by the error-free encoder with $1024$ context channels.}
    \label{fig:res08}
    \vspace{-10pt}
\end{figure*}

\subsection{Environment-Conditioned Downstream Evaluation}
\label{subsec:envaware}

We evaluate the frozen {\sffamily\scshape Channel2World} encoder as an environment-conditioning module for channel matrix-to-position estimation and beam-domain CSI reconstruction. The evaluation uses $16$ BS-centered environments randomly selected from the test factory layouts. For each environment, the channel samples are divided into training and test subsets with an $80$/$20$ split, and all reported metrics are computed on the test subset. We compare the following settings:

$\bullet$ \emph{Test: Site-specific} trains one site-specific model for each selected test environment using only its target-environment training split.

$\bullet$ \emph{Test: Combined} trains one model by pooling the target-environment training splits from all $16$ selected environments.

$\bullet$ \emph{Global} trains one model using all samples from the training environments and no samples from the test environments.

$\bullet$ \emph{Global + fine-tuning} starts from the trained \emph{Global} model and updates a lightweight output module for target-environment adaptation using $128$, $256$, or $1024$ task-labeled samples from the target-environment training split.

$\bullet$ \emph{Proposed} trains two model variants using the same training data as \emph{Global}, conditioned on embeddings produced by the frozen error-aware and error-free {\sffamily\scshape Channel2World} encoders, respectively. At evaluation, each embedding $\mathbf{Z}_{e}$ is inferred from $128$ context channel-position samples from the target-environment training split, without updating the corresponding encoder or downstream model.

$\bullet$ \emph{Mean} and \emph{Shuffled} use the trained \emph{Proposed} model but replace the target-environment embedding during evaluation with either the empirical mean embedding or an embedding inferred from another test environment.

The target-environment reference models (\emph{Test: Site-specific} and \emph{Test: Combined}) and fine-tuning baselines use task-labeled samples from the selected test environments to train or adapt the downstream predictor. In contrast, the proposed method uses target-environment samples only to infer the wireless world embedding and performs no gradient-based target-environment adaptation. The task-specific predictor backbone is kept the same across the target-environment, global, and proposed settings, with the proposed model adding only a lightweight attention-pooling module for $\mathbf{Z}_{e}$. The following results show that the wireless world embedding inferred by {\sffamily\scshape Channel2World} contains downstream-relevant environmental structure that can be exploited in unseen environments without task-specific target-environment adaptation.

\subsubsection{Channel Matrix-to-Position Estimation Results}
\label{subsec:sim_position_results}

The first downstream task, introduced in Section~\ref{subsec:downstream_channel_position}, estimates the BS-centered UE position from a single uplink spatial channel matrix $\bold{H}_{e,i}\in\mathbb{C}^{4\times 8}$ obtained at $7.0$~GHz with a $4\times8$ BS array. Fig.~\ref{fig:res04} shows the position RMSE versus the number of trainable parameters in the downstream model. The proposed environment-conditioned models consistently outperform the unconditioned \emph{Global} model across the tested model sizes. Since the comparison accounts for the parameter increase introduced by the conditioning module, the gain mainly reflects the target-environment information contained in $\mathbf{Z}_{e}$ rather than the larger model size. The proposed models also outperform the \emph{Test: Site-specific} and \emph{Test: Combined} references, although these references use target-environment samples. This trend is attributed to the limited training data available within the selected test environments, which restricts the diversity of channel-position pairs for learning the site-specific mapping. For the fine-tuning baselines in this task, the final linear layer of the trained \emph{Global} model is replaced by an MLP adaptation head with one hidden layer, and only this head is updated using target-environment samples. The fine-tuning baselines improve as more target-environment samples are used. With $128$ samples, fine-tuning remains significantly worse than the proposed conditioning, despite using the same number of target-environment samples as the \emph{Proposed} model. With $1024$ samples, fine-tuning outperforms the proposed model by using substantially more task-labeled target-environment data. The \emph{Mean} and \emph{Shuffled} ablations degrade markedly and perform worse than \emph{Global}, showing that the gain from environment conditioning depends on using an embedding matched to the target environment. The error-aware and error-free encoders provide nearly identical performance, indicating that the considered estimation errors do not critically reduce the usefulness of the inferred embedding for this task.

\subsubsection{Beam-Domain CSI Reconstruction Results}
\label{subsec:sim_csi_reconstruction}

The second downstream task, introduced in Section~\ref{subsec:downstream_beam_reconstruction}, reconstructs the full $8\times16$ DFT beam-domain received-power map from sparse, noisy beam measurements. The evaluation uses an $8\times16$ BS array at $7.0$~GHz with isotropic BS antenna elements and a single isotropic UE antenna. The total BS transmit power, bandwidth, and receiver noise power spectral density are set to $24$~dBm, $20$~MHz, and $-174$~dBm/Hz, respectively. The codebook sizes $|\Omega|=12,16,24,32$ correspond to $2\times6$, $4\times4$, $3\times8$, and $4\times8$ fixed-pattern beam subsets, respectively. The \emph{Test: Site-specific}, \emph{Test: Combined}, \emph{Global}, and fine-tuning models have approximately $1.80$ million parameters, compared with $1.86$ million for the proposed model due to the latent pooling module. For this high-dimensional input-output task, we do not attach a newly initialized adaptation head for the fine-tuning baselines, since training such a head from a small target-environment set may generalize poorly. Instead, fine-tuning starts from the trained \emph{Global} model and updates only its final two-layer MLP prediction block using target-environment samples. Fig.~\ref{fig:res05} reports the reconstruction RMSE in dB. Increasing $|\Omega|$ generally reduces the error, although the trend is not strictly monotonic because different fixed beam-subset patterns provide different levels of information to the learned reconstructor. The proposed models improve over \emph{Global}, although the gain is less noticeable than in position estimation because this task is largely an interpolation of the beam-domain map. The target-environment references reflect the training-data demand of this high-dimensional task. The \emph{Test: Site-specific} model is limited by the amount of per-environment data, while \emph{Test: Combined} benefits from data pooling across environments and becomes more competitive for larger codebooks. The fine-tuning baselines with $128$ and $256$ samples are competitive with the proposed models. However, fine-tuning requires beam-domain CSI labels and additional gradient-based adaptation. The \emph{Mean} and \emph{Shuffled} ablations remain closer to the proposed models than in localization, indicating that reconstruction is less sensitive to the exact environment embedding when sparse beam measurements already provide substantial information for interpolation.

\subsubsection{Environment Reconstruction Results}
\label{subsec:sim_environment_reconstruction}

\begin{comment}
{\renewcommand{\arraystretch}{1.0}
\begin{table}[!t]
	\centering
	\caption{Environment Reconstruction Performance}
	\footnotesize
	\label{tbl02}
	\begin{tabular}{c|c|c}
		\hline
		{\textbf{Model}} & {\textbf{RMSE}} & \textbf{MAE}\\
		\hline \hline
		Error-aware \& 128 context inputs & {\,\,\,3.91~m\,\,\,} & {\,\,\,1.94~m\,\,\,}\\
		\hline
		Error-free \& 1024 context inputs & {\,\,\,3.88~m\,\,\,} &  {\,\,\,1.84~m\,\,\,}\\
		\hline
	\end{tabular} 
	\vspace{-10pt}
\end{table}
}
\end{comment}

The third downstream task, introduced in Section~\ref{subsec:downstream_environment_reconstruction}, evaluates whether the wireless world embedding contains recoverable geometric information. Since the BS-centered AoA specifies only the ray direction, the range to the first BS-side interaction point must be inferred from $\mathbf{Z}_{e}$. We train two range predictors conditioned on embeddings inferred by the error-aware encoder using $128$ context channel-position samples and the error-free encoder using $1024$ context samples, respectively. Despite relying on fewer and perturbed context observations, the error-aware setting achieves an RMSE of $3.91$~m and a mean absolute error (MAE) of $1.94$~m, close to the $3.88$~m RMSE and $1.84$~m MAE obtained in the error-free setting. Fig.~\ref{fig:res08} shows representative reconstruction results, in which the dominant BS-side reflector structures and their approximate spatial extent are recovered. The reconstruction is less accurate near corners, discontinuities, and sparsely observed reflector regions. This limitation arises because the decoder predicts one range value independently for each AoA-defined ray and does not impose surface-continuity constraints. These results indicate that $\mathbf{Z}_{e}$ retains recoverable RF-observable geometry, although the proposed decoder should not be interpreted as a complete environment reconstruction method.

\section{Conclusion}
In this paper, we proposed {\sffamily\scshape Channel2World}, a wireless foundation model that infers BS-centered RF environment representations from MIMO channel-position observations. {\sffamily\scshape Channel2World} aggregates these observations into wireless world embedding tokens and trains the environment encoder through context-query pretraining. Evaluation using ray-tracing channel data from 26,000 randomized BS-centered environments demonstrated that the learned embeddings capture site-specific propagation structures and generalize to unseen building layouts. The inferred wireless world embeddings improved performance on downstream tasks, including UE position estimation and sparse beam-domain CSI reconstruction, compared with unconditioned and mismatched-latent baselines. In addition, the RF-observable geometry reconstruction task further showed that the embeddings contain sufficient information to recover dominant reflector structures. These results suggest that channel-position observations can provide reusable RF environment priors for environment-aware AI-native wireless systems. Future work includes validating the proposed framework on measured multi-site channel datasets.

\bibliographystyle{IEEEtran}
\bibliography{HJM_ref}

\end{document}